\documentclass[journal]{IEEEtran}

\usepackage{cite}
\usepackage{graphicx}
\usepackage{amsmath} 
\usepackage{multirow}
\usepackage{amssymb}
\usepackage{url}
\usepackage{booktabs}

\usepackage{pdflscape}

\makeatletter
\let\MYcaption\@makecaption
\makeatother

\usepackage[font=footnotesize]{subcaption}

\makeatletter
\let\@makecaption\MYcaption
\makeatother

\begin{document}


\title{Quantitative Test of the Evolution of Geant4 Electron Backscattering Simulation }

\author{Tullio Basaglia, Min Cheol Han, Gabriela Hoff, Chan Hyeong Kim, Sung Hun Kim,  Maria Grazia Pia, and Paolo Saracco 
\thanks{Manuscript received June 29, 2016.}
\thanks{T. Basaglia is with 
	CERN, CH1211 Gen\`eve 23, Switzerland
	(e-mail: Tullio.Basaglia@cern.ch).}
\thanks{G. Hoff is with Instituto Polit\'ecnico - IPRJ, Universidade do Estado do Rio de Janeiro, Nova Friburgo - RJ,  Brazil
	(e-mail: ghoff.gesic@gmail.com).}
\thanks{ S. H. Kim,  M. C. Han and C. H. Kim are with 
	the Department of Nuclear Engineering, Hanyang University, 
        Seoul 133-791, Korea 
	(e-mail: ksh4249@hanyang.ac.kr, mchan@hanyang.ac.kr, chkim@hanyang.ac.kr).}
\thanks{M. G. Pia and P. Saracco are with 
	INFN Sezione di Genova, Via Dodecaneso 33, I-16146 Genova, Italy 
	(phone: +39 010 3536328, fax: +39 010 313358,
	e-mail: MariaGrazia.Pia@ge.infn.it, Paolo.Saracco@ge.infn.it).}
}

\maketitle

\begin{abstract}
Evolutions of Geant4 code have affected the simulation
of electron backscattering with respect to previously published results.
Their effects are quantified by analyzing the compatibility of the simulated
electron backscattering fraction with a large collection of experimental data
for a wide set of physics configuration options available in Geant4.
Special emphasis is placed on two electron scattering implementations first
released in Geant4 version 10.2: the Goudsmit-Saunderson multiple scattering
model and a single Coulomb scattering model based on Mott cross section
calculation. The new Goudsmit-Saunderson multiple scattering model appears to
perform equally or less accurately than the model implemented in previous Geant4
versions, depending on the electron energy. The new Coulomb scattering model was
flawed from a physics point of view, but computationally fast in Geant4 version 10.2; the physics
correction released in Geant4 version 10.2p01 severely degrades its
computational performance.
Problems observed
in electron backscattering simulation in previous publications have been 
addressed by evolutions in the Geant4 geometry domain.
\end{abstract}

\begin{IEEEkeywords}
Monte Carlo, simulation, Geant4, electrons
\end{IEEEkeywords}

\section{Introduction}
\label{sec_intro}

\IEEEPARstart{T}{he} simulation of electron backscattering based on Geant4
\cite{g4nim,g4tns} has been investigated in a variety of configurations
\cite{tns_ebscatter1, tns_ebscatter2}, as this observable is a sensitive probe
of multiple and single scattering modeling in Monte Carlo codes for particle
transport.

Apparent 
anomalies, leading to the suppression of backscattering,
were observed in association with some physics configurations
\cite{tns_ebscatter1}; dedicated investigations \cite{tns_ebscatter2} hinted
that some of the step limitation algorithms related to the treatment of multiple
scattering could be sensitive to peculiarities of the Geant4 geometry domain, while
single scattering simulation appeared to be immune from such effects.
The observed inconsistency of the backscattering simulation outcome, which appeared to depend on the configuration of the 
experimental setup in the simulation, precluded unequivocal quantification of the accuracy of
Geant4 multiple and single scattering models and their relative comparison in
\cite{tns_ebscatter2}.
These issues have been addressed in Geant4 10.1p02 and later versions, thus
allowing consistent quantification of Geant4 capability to simulate electron
backscattering in a variety of physics configurations and of their relative ability 
to reproduce experimental measurements.

Geant4 version 10.2 also includes two new electron scattering implementations
and a new predefined \textit{PhysicsConstructor}, which are quantitatively
evaluated in an experimental test case for the first time in this paper.
The Goudsmit-Saunderson \cite{goudsmit1,goudsmit2} multiple
scattering model  was completely reimplemented in Geant4 10.2 on the basis of a
reworked analytical foundation of the code: although the new model retains the same
class name present in previous Geant4 versions, the code is entirely new.
A predefined \textit{PhysicsConstructor} using the Goudsmit-Saunderson model was
released for the first time in Geant4 10.2.
A single scattering model based on the Mott cross section \cite{mott1929,
mott1932, boschini_2013}, which did not work properly in previous Geant4
versions, and for this reason could not be used in the tests of
\cite{tns_ebscatter1, tns_ebscatter2}, was modified to become functional in
Geant4 10.2.
This model extends the provision of methods to simulate electron scattering as a
discrete process in Geant4, as an alternative to condensed history schemes 
usually adopted in particle transport through multiple scattering modeling.


This paper documents quantitatively the effects of these Geant4 evolutions on the
fraction of backscattered electrons.
This observable is the most basic probe of the simulation of electron scattering
in particle transport codes;
its assessment is preparative
to the validation of more complex observables
related to electron scattering, such as the energy and angular distributions of
scattered particles.
Special emphasis is given in the following section to the characterization of
the new physics modeling options available in Geant4 with respect to
experimental measurements, and the assessment of their capabilities in
comparison to other, previously available, options.
All the results are based on statistical data analysis methods to ensure their
objectiveness.




\tabcolsep=3pt
\begin{table*}[htbp]
  \centering
  \caption{Multiple and single scattering settings in the simulation configurations evaluated in this test}
    \begin{tabular}{llllclc}
    \toprule
    {\bf Configuration} & {\bf Description} 		& {\bf Process class} 			& {\bf Model class} 	& {\bf Version}	& {\bf Step Limitation} & {\bf RangeFactor} \\
    \midrule
    \textbf{Urban} 	& Urban model, user step limit  	 & G4eMultipleScattering & G4UrbanMscModel		   &      & Safety \textit{(default)} & 0.04 \textit{(default)} \\
   \textbf{UrbanBRF} 	& Urban model      			 & G4eMultipleScattering & G4UrbanMscModel 		   &      & DistanceToBoundary        & 0.01                    \\
   \textbf{WentzelBRF} 	& WentzelVI model 			 & G4eMultipleScattering & G4WentzelVIModel 		   &      & DistanceToBoundary        & 0.01                    \\
  \textbf{WentzelBRFP} 	& WentzelVI model, $\theta_{limit}$=0.15  & G4eMultipleScattering & G4WentzelVIModel 		   &      & DistanceToBoundary        & 0.01                    \\
   \textbf{GS} 		& Goudsmit-Saunderson                    & G4eMultipleScattering & G4GoudsmitSaundersonModel       & 10.1 & Safety 		      & 0.01                    \\
			& Goudsmit-Saunderson, Moli\`ere         & G4eMultipleScattering & G4GoudsmitSaundersonModel       & 10.2 & SafetyPlus  	      & 0.12                    \\
   \textbf{GSBRF} 	& Goudsmit-Saunderson                    & G4eMultipleScattering & G4GoudsmitSaundersonModel       & 10.1 & DistanceToBoundary        & 0.01                    \\
			& Goudsmit-Saunderson, Moli\`ere 	 & G4eMultipleScattering & G4GoudsmitSaundersonModel       & 10.2 & DistanceToBoundary        & 0.12                    \\
   \textbf{GSBRF1} 	& Goudsmit-Saunderson  			 & G4eMultipleScattering & G4GoudsmitSaundersonModel       & 10.2 & DistanceToBoundary        & 0.01                    \\
   \textbf{GSERF} 	& Goudsmit-Saunderson, Moli\`ere 	 & G4eMultipleScattering & G4GoudsmitSaundersonModel       & 10.2 & Safety 	 	      & 0.12                    \\
   \textbf{GSPWA} 	& Goudsmit-Saunderson, PWA  		 & G4eMultipleScattering & G4GoudsmitSaundersonModel       & 10.2 & SafetyPlus  	      & 0.12                    \\
   \textbf{GSPWABRF} 	& Goudsmit-Saunderson, PWA  		 & G4eMultipleScattering & G4GoudsmitSaundersonModel       & 10.2 & DistanceToBoundary        & 0.12                    \\
   \textbf{GSPWAERF} 	& Goudsmit-Saunderson, PWA  		 & G4eMultipleScattering & G4GoudsmitSaundersonModel       & 10.2 & Safety  		      & 0.12                    \\
   \textbf{Coulomb} 	& Single scattering      		 & G4CoulombScattering   & G4eCoulombScatteringModel       &      &                           &                         \\
   \textbf{CoulombMott} & Single scattering, Mott    		 & G4CoulombScattering   & G4eSingleCoulombScatteringModel & 10.2 &                           &                         \\

\bottomrule
    \end{tabular}%
  \label{tab_msconf}%
\end{table*}

\tabcolsep=6pt

\begin{table*}[htbp]
  \centering
  \caption{Predefined Geant4 Electromagnetic PhysicsConstructors evaluated in this test }
    \begin{tabular}{llclcc}
    \toprule
    {\bf Configuration} 	& {\bf PhysicsConstructor class} 		& {\bf Version} & {\bf Step Limitation} 	& {\bf  RangeFactor}  & {\bf ThetaLimit}\\
    \midrule
    \textbf{EmLivermore} & G4EmLivermorePhysics 	& 10.1             & DistanceToBoundary & 0.01 &      \\
			 &                              & 10.2             & DistanceToBoundary & 0.02 &      \\
    \textbf{EmStd} 	 & G4EmStandardPhysics		&                  & Safety 		& 0.04 &      \\
    \textbf{EmOpt1} 	 & G4EmStandardPhysics\_option1 & 10.1             & Minimal 		& 0.04 &      \\
			 &                              & 10.2             & Minimal		& 0.20 &      \\
    \textbf{EmOpt2} 	 & G4EmStandardPhysics\_option2 & 10.1             & Minimal 		& 0.04 &      \\
			 &                              & 10.2             & Minimal		& 0.20 &      \\
    \textbf{EmOpt3} 	 & G4EmStandardPhysics\_option3 &                  & DistanceToBoundary & 0.04 &      \\
    \textbf{EmOpt4} 	 & G4EmStandardPhysics\_option4 & 10.1             & SafetyPlus  	& 0.02 &      \\
			 &                              & 10.1p02, 10.1p03 & DistanceToBoundary & 0.02 &      \\
			 &                              & 10.2             & DistanceToBoundary & 0.02 &      \\
    \textbf{EmSS} 	 & G4EmStandardPhysicsSS	&                  & Safety	 	& 0.04 & 0    \\
    \textbf{EmWVI} 	 & G4EmStandardPhysicsWVI 	& 10.1             & Safety		& 0.04 & 0.15 \\		
			 &                              & 10.2             & Safety		& 0.04 & 0.02 \\
    \textbf{EmGS} 	 & G4EmStandardPhysicsGS 	& 10.2             & SafetyPlus		& 0.12 &      \\

\bottomrule
    \end{tabular}%
  \label{tab_pcconf}%
\end{table*}

\section{Simulation and Data Analysis Features}

\subsection{Simulation configuration}
\label{sec_config}

The validation tests documented in this paper concern Geant4 versions from
10.1p02 to 10.2p02, which were released after the analyses reported in
\cite{tns_ebscatter1} and \cite{tns_ebscatter2}.
It is worthwhile to note that 
Geant4 version 10.1p03 is more recent than version 10.2.
The documentation of the performance of a variety of Geant4 versions is relevant
to the experimental community, where several versions of Geant4 are in use at
the same time in different experiments, whose simulation production strategies
do not always coincide with the rapid turnover of Geant4 version releases.
It also provides significant information for the improvement of the Geant4 software 
development process and, more generally, for software engineering measurements
relevant to large scale systems \cite{ronchieri_chep2015}.

The experimental scenario and simulation execution environment
pertinent to this paper are the same as in \cite{tns_ebscatter1}, where they are
extensively described; interested readers can find detailed information in 
\cite{tns_ebscatter1} and in the references cited therein.
The additional details given below concern simulation features pertinent to
Geant4 versions 10.1p02 to 10.2p02.

The physics configurations considered in this validation test are summarized in
Tables \ref{tab_msconf} and \ref{tab_pcconf}, which concern user-defined
and predefined physics settings, respectively.
Electron and photon interactions other than electron multiple and single
scattering are based on the EEDL \cite{eedl} and EPDL \cite{epdl97} data
libraries \cite{lowe_e, lowe_chep, lowe_nss} in the configurations listed in
Table \ref{tab_msconf}, while
in the configurations of Table \ref{tab_pcconf} they are determined
by the predefined physics settings pertinent to each PhysicsConstructor.

Predefined electromagnetic PhysicsConstructors in general use a combination of 
different electron scattering models and differ in the configuration of 
other electron and photon interactions; therefore it is difficult to ascertain
the contribution of each physics modeling component to the accuracy of the 
simulated observable.
Although some of them (\textit{G4EmStandardPhysicsGS}, \textit{G4EmStandardPhysicsSS}
\textit{G4EmStandardPhysicsWVI}) are defined as ``experimental physics'' in 
Geant4 user documentation associated with Geant4 version 10.2 
\cite{g4appldevguide},
these PhysicsConstructors use theoretical models to describe electron and photon 
interactions with matter.

The user-defined physics configurations listed in Table \ref{tab_msconf} are
intended to facilitate investigation of the effects of a multiple or single scattering model,
with specified settings of options and parameters, on the simulated observable.
For this purpose they instantiate a unique electron scattering process (multiple
or single), with a unique model associated to it, and use the same configuration 
of electron-photon interactions other than electron scattering to highlight the 
effects specific to electron scattering modeling.



The new Goudsmit-Saunderson multiple scattering model first released in Geant4 10.2 can
calculate the screening parameter according to the Moli\`ere formula (by default) or
using elastic cross sections deriving from partial wave analysis calculations: the
corresponding configurations are identified in Table \ref{tab_msconf} as ``GS''
and ``GSPWA'', respectively.
Additionally, it provides three step limitation algorithm options, identified as
\textit{SafetyPlus}, \textit{DistanceToBoundary} and \textit{Safety}.
The \textit{SafetyPlus} algorithm associated with the reimplemented
Goudsmit-Saunderson model corresponds to the \textit{Safety} step limitation
algorithm associated with the Urban \cite{urban, urban2006} multiple scattering
model, and is used by default; the \textit{DistanceToBoundary} algorithm
corresponds to the same option of the Urban model; the \textit{Safety} algorithm
corresponds to EGSnrc \cite{egsnrc} error-free stepping algorithm.
Therefore, six configurations related to the reimplemented Goudsmit-Saunderson
model are considered in the validation analysis, which reflect the combination
of options for the calculation of the screening parameter and for the step
limitation algorithm; they are listed in Table \ref{tab_msconf}.
Additionally, a configuration identified as ``GSBRF1'' is included in the test
for the purpose of studying the effect of the \textit{RangeFactor} parameter:
it is identical to ``GSBRF'', except for this different setting.

A predefined PhysicsConstructor \textit{G4EmStandardPhysicsGS},
which uses the reimplemented Goudsmit-Saunderson multiple scattering model, was
first introduced in Geant4 10.2 and is listed in Table \ref{tab_pcconf} along
with other previously released predefined PhysicsConstructors.
In this class the Goudsmit-Saunderson model is configured with its default 
 Moli\`ere screening and \textit{SafetyPlus} step limitation options; additionally,
the \textit{RangeFactor} parameter is set to 0.12.


The new single scattering model released in Geant4 10.2 is associated with the
\textit{G4eSingleCoulombScatteringModel} class.
The physics configuration which uses it is identified in this paper as
``CoulombMott''.
According to the Geant4 software documentation, this model is
applicable to electrons of energy greater than 200~keV, incident on medium-light
target nuclei \cite{g4physmanual}.

Nevertheless, simulations involving this model were executed over all the
experimental test cases considered in this paper, which span the range of atomic
numbers from 4 to 92 and involve electron beam energies both below and above
200~keV, to characterize the behaviour of this model also outside its nominal
applicability.
This extended investigation is intended to quantify more objectively the
capabilities of this model, considering that the qualitative definition of
medium-light nuclei is susceptible to subjective interpretation and that
electrons that are originally incident on the target with energy greater than
the documented 200~keV threshold of applicability lose energy in the course of
the transport due to interactions with the traversed medium.
The results of using \textit{G4eSingleCoulombScatteringModel} outside its nominal
range of applicability are discussed in section \ref{sec_mott}.
It is worthwhile to note that no warnings of improper use of this model were
issued when it was invoked outside its nominal domain of applicability, nor were any
exceptions thrown in the course of the execution of the simulations.

%


\subsection{Data Analysis}
\label{sec_analysis}

The validation test concerns the fraction of electrons that are backscattered
from a semi-infinite or infinite target of pure elemental composition.
The reference experimental data involved in the validation process are the same
as in \cite{tns_ebscatter1};
for all aspects related to experimental data the reader is invited to consult reference \cite{tns_ebscatter1}.

Compatibility between simulation and experiment, as well as differences in
compatibility with experiment associated with different simulation
configurations, are assessed by means of statistical methods that are
described in detail in \cite{tns_ebscatter1} and \cite{tns_ebscatter2}.
The Statistical Toolkit \cite{gof1, gof2} and R \cite{R} are used as software
instruments for data analysis.

For convenience, compatibility with experimental data is
summarized by a variable defined as ``efficiency'', which represents the
fraction of test cases where the p-value resulting from goodness-of-fit tests is
larger than the predefined significance level.
The uncertainties on the efficiencies are calculated according to a method
based on Bayes' theorem \cite{paterno_2004}, which delivers meaningful results
also in limiting cases, i.e. for efficiencies very close to 0 or to 1, where the
conventional method based on the binomial distribution \cite{frodesen}
produces unreasonable values.
Apart from these special cases, both methods deliver identical results within
the number of significant digits reported in the following tables.

As discussed in \cite{tns_ebscatter1}, the results reported in section
\ref{sec_results} are based on the Anderson-Darling
\cite{anderson1952,anderson1954} goodness-of-fit test, since the outcome of
other tests (Cramer-von Mises \cite{cramer,vonmises}, Kolmogorov-Smirnov
\cite{kolmogorov1933,smirnov} and Watson \cite{watson}) is statistically
equivalent.

Contingency tables are used for categorical data analysis, where simulation
configurations represent categories.
They are based on the results of the Anderson-Darling test; their entries 
count the number of test cases associated with a given simulation
configuration for which the null hypothesis of compatibility between simulated
and backscattered data is rejected or fails to be rejected.
It is worth stressing that what is compared in contingency tables is the
capability of the simulation configurations subject to test to produce a
fraction of backscattered electrons statistically compatible with experiment,
not the backscattering fraction simulated with different configurations.

A variety of tests is applied to mitigate the risk of systematic effects in
categorical data analysis: Fisher's exact test \cite{fisher}, Barnard's exact test
\cite{barnard} using the Z-pooled statistic \cite{suissa} and the CSM
approximation, Boschloo's exact test \cite{boschloo} and Pearson's $\chi^2$ \cite{pearson}
test, when the entries in the cells of a table are consistent with its
applicability.
The power of tests for categorical data analysis is not
well established yet; Boschloo's test and Barnard's exact test calculated using the Z-pooled statistic
are deemed more powerful than Fisher's exact test for the analysis of 2x2
contingency tables \cite{agresti_1992, andres_1994,andres_2004}.

The significance level is set at 0.01 both for goodness-of-fit tests and for the
analysis of contingency tables.


\subsection{Computational performance}

The intrinsic characteristics of the simulations, which reproduce different
experimental models in terms of target shape and size and are executed in a
heterogeneous computational environment \cite{tns_ebscatter1}, allow only a
qualitative appraisal of the computational performance associated with the
various physics configurations considered in this paper.
Nevertheless, this complementary information provides valuable guidance for 
practical use of the physics configurations examined in this paper in realistic
experimental scenarios; it is therefore discussed in the following sections
that document the results derived from the various simulation configurations.

The variability of the computational environment and of the experimental
scenarios involved in the test is partly taken into account by
rescaling the CPU (central processing unit) time spent for each simulation
according to the hardware characteristics of the node where it was executed,
and by evaluating the computational performance in relative terms, i.e. with
respect to a configuration taken as a common reference for comparison
(the ``EmStd'' configuration, unless otherwise specified).

A smaller number of events was generated in some simulations using
\textit{G4eSingleCoulombScatteringModel} than in the other simulation
configurations due to the exceedingly slow computational performance
associated with this model,
which would have required an unsustainable amount of computing resources
to produce a simulated data sample of the same size as in the other test cases.
These smaller data samples are reflected in larger error bars appearing in some
plots concerning \textit{G4eSingleCoulombScatteringModel}.



\tabcolsep=2pt
\begin{table*}[htbp]
  \centering
  \caption{Efficiency calculated for different simulation configurations and Geant4 versions, in three energy ranges}
    \begin{tabular}{lcccccccccccc}
    \toprule
				& \multicolumn{4}{c}{1-20 keV}  & \multicolumn{4}{c}{20-100 keV}        & \multicolumn{4}{c}{$\ge$100 keV}  \\
\cmidrule(lr) {2-5}
\cmidrule(lr) {6-9}
\cmidrule(lr) {10-13}
    \textit{Configuration} & \textbf{10.1p03} & \textbf{10.2} & \textbf{10.2p01} & \textbf{10.2p02} & \textbf{10.1p03} & \textbf{10.2} & \textbf{10.2p01} & \textbf{10.2p02}& \textbf{10.1p03} & \textbf{10.2} & \textbf{10.2p01} & \textbf{10.2p02}\\
    \midrule
    \textbf{Urban} 	   & 0.08$\pm$0.02    & 0.16$\pm$0.03 & 0.16$\pm$0.03 & 0.16$\pm$0.03   & 0.22$\pm$0.04    & 0.25$\pm$0.04 & 0.25$\pm$0.04 & 0.21$\pm$0.04   & 0.63$\pm$0.06    & 0.64$\pm$0.06 & 0.64$\pm$0.06 & 0.64$\pm$0.06    \\
    \textbf{UrbanBRF} 	   & 0.16$\pm$0.03    & 0.10$\pm$0.03 & 0.16$\pm$0.03 & 0.16$\pm$0.03   & 0.33$\pm$0.04    & 0.29$\pm$0.04 & 0.38$\pm$0.04 & 0.39$\pm$0.04   & 0.68$\pm$0.06    & 0.63$\pm$0.06 & 0.63$\pm$0.06 & 0.63$\pm$0.06    \\
    \textbf{GS} 	   & 0.51$\pm$0.04    & 0.57$\pm$0.04 & 0.56$\pm$0.04 & 0.56$\pm$0.04   & 0.32$\pm$0.04    & 0.18$\pm$0.04 & 0.16$\pm$0.04 & 0.16$\pm$0.04   & 0.77$\pm$0.06    & 0.52$\pm$0.07 & 0.55$\pm$0.06 & 0.55$\pm$0.06    \\
     \textbf{GSBRF} 	   & 0.37$\pm$0.04    & 0.38$\pm$0.04 & 0.57$\pm$0.04 & 0.57$\pm$0.04   & 0.51$\pm$0.05    & 0.46$\pm$0.05 & 0.14$\pm$0.05 & 0.14$\pm$0.05   & 0.95$\pm$0.03    & 0.61$\pm$0.06 & 0.55$\pm$0.06 & 0.55$\pm$0.06    \\
    \textbf{GSBRF1} 	   &                  & 0.35$\pm$0.04 & 0.42$\pm$0.04 & 0.42$\pm$0.04   &                  & 0.45$\pm$0.05 & 0.44$\pm$0.05 & 0.45$\pm$0.05   &                  & 0.61$\pm$0.06 & 0.59$\pm$0.06 & 0.57$\pm$0.06    \\
    \textbf{GSERF} 	   &                  & 0.32$\pm$0.04 & 0.31$\pm$0.04 & 0.32$\pm$0.04   &                  & 0.49$\pm$0.05 & 0.43$\pm$0.05 & 0.46$\pm$0.05   &                  & 0.63$\pm$0.06 & 0.61$\pm$0.06 & 0.59$\pm$0.06    \\
    \textbf{GSPWA} 	   &                  & 0.15$\pm$0.03 & 0.15$\pm$0.03 & 0.15$\pm$0.03   &                  & 0.17$\pm$0.04 & 0.15$\pm$0.04 & 0.15$\pm$0.04   &                  & 0.61$\pm$0.06 & 0.61$\pm$0.06 & 0.61$\pm$0.06    \\
    \textbf{GSPWABRF}      &                  & 0.39$\pm$0.04 & 0.16$\pm$0.04 & 0.16$\pm$0.04   &                  & 0.44$\pm$0.05 & 0.11$\pm$0.05 & 0.11$\pm$0.05   &                  & 0.57$\pm$0.06 & 0.55$\pm$0.06 & 0.55$\pm$0.06    \\
    \textbf{GSPWAERF}      &                  & 0.44$\pm$0.04 & 0.44$\pm$0.04 & 0.44$\pm$0.04   &                  & 0.51$\pm$0.04 & 0.46$\pm$0.04 & 0.47$\pm$0.04   &                  & 0.52$\pm$0.07 & 0.55$\pm$0.07 & 0.55$\pm$0.07    \\
    \textbf{WentzelBRF}    & 0.27$\pm$0.03    & 0.27$\pm$0.04 & 0.27$\pm$0.04 & 0.27$\pm$0.04   & 0.17$\pm$0.04    & 0.18$\pm$0.04 & 0.18$\pm$0.04 & 0.19$\pm$0.04   & 0.80$\pm$0.05    & 0.82$\pm$0.05 & 0.82$\pm$0.05 & 0.82$\pm$0.05    \\
    \textbf{WentzelBRFP}   & 0.47$\pm$0.04    & 0.35$\pm$0.04 & 0.35$\pm$0.04 & 0.35$\pm$0.04   & 0.45$\pm$0.04    & 0.44$\pm$0.04 & 0.43$\pm$0.04 & 0.43$\pm$0.04   & 0.86$\pm$0.05    & 0.86$\pm$0.05 & 0.84$\pm$0.05 & 0.84$\pm$0.05    \\
    \textbf{Coulomb} 	   & 0.49$\pm$0.04    & 0.36$\pm$0.04 & 0.36$\pm$0.04 & 0.37$\pm$0.04   & 0.46$\pm$0.05    & 0.46$\pm$0.05 & 0.46$\pm$0.05 & 0.46$\pm$0.05   & 0.80$\pm$0.05    & 0.79$\pm$0.05 & 0.79$\pm$0.05 & 0.80$\pm$0.05    \\
    \textbf{CoulombMott}   &                  & $<$0.01       & $<$0.01       & $<$0.01         &                  & $<$0.01       & 0.16$\pm$0.03 & 0.16$\pm$0.03   &                  & $<$0.02       & 0.96$\pm$0.03 & 0.96$\pm$0.03    \\
    \textbf{EmLivermore}   & 0.13$\pm$0.03    & 0.11$\pm$0.03 & 0.11$\pm$0.03 & 0.11$\pm$0.03   & 0.29$\pm$0.04    & 0.24$\pm$0.04 & 0.24$\pm$0.04 & 0.25$\pm$0.04   & 0.61$\pm$0.06    & 0.61$\pm$0.06 & 0.61$\pm$0.06 & 0.61$\pm$0.06    \\
    \textbf{EmStd} 	   & 0.13$\pm$0.03    & 0.08$\pm$0.03 & 0.08$\pm$0.03 & 0.08$\pm$0.03   & 0.19$\pm$0.04    & 0.16$\pm$0.03 & 0.16$\pm$0.03 & 0.16$\pm$0.03   & 0.71$\pm$0.06    & 0.86$\pm$0.05 & 0.86$\pm$0.05 & 0.86$\pm$0.05    \\
    \textbf{EmOpt1} 	   & $<$0.01  	      & $<$0.01       & $<$0.01       & $<$0.01         & $<$0.01  	   & $<$0.01       & $<$0.01       & $<$0.01         & 0.41$\pm$0.06    & 0.39$\pm$0.06 & 0.39$\pm$0.06 & 0.39$\pm$0.06    \\
    \textbf{EmOpt2} 	   & $<$0.01  	      & $<$0.01       & $<$0.01       & $<$0.01         & $<$0.01  	   & $<$0.01       & $<$0.01       & $<$0.01         & 0.41$\pm$0.06    & 0.39$\pm$0.06 & 0.39$\pm$0.06 & 0.39$\pm$0.06    \\
    \textbf{EmOpt3} 	   & 0.21$\pm$0.03    & 0.17$\pm$0.03 & 0.17$\pm$0.03 & 0.17$\pm$0.03   & 0.14$\pm$0.03    & 0.19$\pm$0.04 & 0.19$\pm$0.04 & 0.19$\pm$0.04   & 0.68$\pm$0.06    & 0.75$\pm$0.06 & 0.75$\pm$0.06 & 0.75$\pm$0.06    \\
    \textbf{EmOpt4} 	   & 0.23$\pm$0.03    & 0.10$\pm$0.03 & 0.10$\pm$0.03 & 0.10$\pm$0.03   & 0.21$\pm$0.04    & 0.24$\pm$0.04 & 0.24$\pm$0.04 & 0.24$\pm$0.04   & 0.73$\pm$0.06    & 0.66$\pm$0.06 & 0.66$\pm$0.06 & 0.68$\pm$0.06    \\
    \textbf{EmWVI} 	   & 0.45$\pm$0.04    & 0.36$\pm$0.04 & 0.36$\pm$0.04 & 0.36$\pm$0.04   & 0.46$\pm$0.05    & 0.47$\pm$0.05 & 0.47$\pm$0.05 & 0.47$\pm$0.05   & 0.82$\pm$0.05    & 0.82$\pm$0.05 & 0.82$\pm$0.05 & 0.82$\pm$0.05    \\
    \textbf{EmSS} 	   & 0.46$\pm$0.04    & 0.35$\pm$0.04 & 0.35$\pm$0.04 & 0.35$\pm$0.04   & 0.51$\pm$0.05    & 0.53$\pm$0.05 & 0.53$\pm$0.05 & 0.53$\pm$0.05   & 0.82$\pm$0.05    & 0.82$\pm$0.05 & 0.82$\pm$0.05 & 0.82$\pm$0.05    \\
    \textbf{EmGS} 	   &                  & 0.58$\pm$0.04 & 0.58$\pm$0.04 & 0.58$\pm$0.04   &                  & 0.18$\pm$0.04 & 0.19$\pm$0.04 & 0.19$\pm$0.04   &                  & 0.54$\pm$0.06 & 0.50$\pm$0.06 & 0.50$\pm$0.06    \\
    \bottomrule
    \end{tabular}%
  \label{tab_eff}%
\end{table*}%
\tabcolsep=6pt


\section{Results}

\label{sec_results}

\tabcolsep=2pt
\begin{table*}[htbp]
  \centering
  \caption{P-values deriving from the analysis of contingency tables that compare the
compatibility with experiment obtained with different configuration options of the Goudsmit-Saunderson model in Geant4 10.2p02}
    \begin{tabular}{lrrrrrrrrrrrrrrr}
    \toprule
           & \multicolumn{5}{c}{GSBRF $<$20 keV }  & \multicolumn{5}{c}{GSPWAERF 20-100 keV  } & \multicolumn{5}{c}{GSPWA $\ge$100 keV  } \\
\cmidrule(lr){2-6}
\cmidrule(lr){7-11}
\cmidrule(lr){12-16}
    \multicolumn{1}{l}{Configuration} & \textit{Fisher} & $\chi^2$ & \textit{Z-pooled} & \textit{Boschloo} & \textit{CSM}  & \textit{Fisher} & $\chi^2$ & \textit{Z-pooled} & \textit{Boschloo} & \textit{CSM}  & \textit{Fisher} & $\chi^2$ & \textit{Z-pooled} & \textit{Boschloo} & \textit{CSM} \\
\midrule
    GS       & 0.903     & 0.807     & 0.873     & 0.850     & 0.763     & $< 0.001$ & $< 0.001$ & $< 0.001$ & $< 0.001$ & $< 0.001$ & 0.702 & 0.566 & 0.681 & 0.615 & 0.498 \\
    GSBRF    &           &           &           &           &           & $< 0.001$ & $< 0.001$ & $< 0.001$ & $< 0.001$ & $< 0.001$ & 0.702 & 0.566 & 0.681 & 0.615 & 0.498 \\
    GSBRF1   & 0.015     & 0.011     & 0.013     & 0.012     & 0.022     & 0.789     & 0.688     & 0.753     & 0.731     & 0.750     & 1.000 & 0.847 & 0.917 & 1.000 & 0.737 \\
    GSERF    & $< 0.001$ & $< 0.001$ & $< 0.001$ & $< 0.001$ & $< 0.001$ & 0.893     & 0.789     & 0.853     & 0.836     & 0.984     & 1.000 & 0.847 & 0.917 & 1.000 & 0.737 \\
    GSPWA    & $< 0.001$ & $< 0.001$ & $< 0.001$ & $< 0.001$ & $< 0.001$ & $< 0.001$ & $< 0.001$ & $< 0.001$ & $< 0.001$ & $< 0.001$ &       &       &       &       &       \\
    GSPWABRF & $< 0.001$ & $< 0.001$ & $< 0.001$ & $< 0.001$ & $< 0.001$ & $< 0.001$ & $< 0.001$ & $< 0.001$ & $< 0.001$ & $< 0.001$ & 0.702 & 0.566 & 0.681 & 0.615 & 0.498 \\
    GSPWAERF & 0.039     & 0.029     & 0.033     & 0.033     & 0.058     &           &           &           &           &           & 0.702 & 0.566 & 0.681 & 0.615 & 0.498 \\
\midrule
    EmGS     & 1.000     & 0.902     & 0.949     & 1.000     & 0.861     & $< 0.001$ & $< 0.001$ & $< 0.001$ & $< 0.001$ & $< 0.001$ & 0.342 & 0.254 & 0.280 & 0.280 & 0.277 \\
    \bottomrule
    \end{tabular}%
  \label{tab_GS1022}%
\end{table*}%
\tabcolsep=6pt

\begin{table}[htbp]
  \centering
  \caption{Efficiency of Goudsmit-Saunderson configurations with RangeFactor-0.10 in a simulation environment based on Geant4 10.2p01 }
    \begin{tabular}{lrrr}
    \toprule
    Option & $<$20 keV & 20-100 keV & $\ge$100 keV \\
    \midrule
    GS       & 0.65 $\pm$ 0.04 & 0.21 $\pm$ 0.04 & 0.54 $\pm$ 0.06 \\
    GSBRF    & 0.62 $\pm$ 0.04 & 0.20 $\pm$ 0.04 & 0.54 $\pm$ 0.06 \\
    GSERF    & 0.37 $\pm$ 0.04 & 0.47 $\pm$ 0.04 & 0.57 $\pm$ 0.06 \\
    GSPWA    & 0.19 $\pm$ 0.03 & 0.21 $\pm$ 0.04 & 0.64 $\pm$ 0.06 \\
    GSPWABRF & 0.18 $\pm$ 0.04 & 0.15 $\pm$ 0.03 & 0.63 $\pm$ 0.06 \\
    GSPWAERF & 0.43 $\pm$ 0.04 & 0.47 $\pm$ 0.05 & 0.59 $\pm$ 0.06 \\
    \bottomrule
    \end{tabular}%
  \label{tab_effGS010}%
\end{table}%

\subsection{General overview}


The efficiencies resulting from the Anderson-Darling goodness-of-fit test are
reported in Table \ref{tab_eff} for all the simulation configurations considered
in this validation test;
they are listed for Geant4 versions 10.1p03, 10.2, 10.2p01 and 10.2p02. 
The efficiencies for Geant4 10.1p02 are the same as for 10.1p03 within the
number of significant digits appearing in Table \ref{tab_eff};
those for 10.2p01 and 10.2p02 appear very similar.
The outcome of the validation tests is discussed in detail in the following
sections, with emphasis on the new physics models first introduced in Geant4
version 10.2 and the evolution with respect to previously published results.

Experimental uncertainties are reported in the plots discussed in the following
sections when they are documented in the corresponding publications; 
statistical uncertainties of the simulated data that are smaller than the marker
size are not visible in the figures.


\subsection{Dependencies Between Physics and Geometry Domains}

Corrections to the Geant4 geometry domain were included in Geant4 10.1p02
to address the issues of apparent interplay with some multiple scattering 
simulation features described in \cite{tns_ebscatter2}.
These corrections fix the problem in Geant4 10.1p02 when the target and the
detection hemisphere are adjacent (i.e. they share the boundary surface), but
anomalies are still noticeable in simulations based on that version when the
two geometrical components of the experimental model are displaced.
This issue was addressed by further corrections implemented in Geant4 10.2 
and later released also in Geant4 10.1p03.
These two sets of corrections solve the problems described in
\cite{tns_ebscatter2}.

The results reported in this paper were produced with adjacent target and
detector hemisphere; they are exempt from the previously mentioned problems
related to the geometrical configuration.

\subsection{Goudsmit-Saunderson Multiple Scattering Model}
\label{sec_GS}

The data analysis concerning the Goudsmit-Saunderson multiple scattering
addresses a few distinct issues: evaluating whether the options for the
screening parameter and step limitation significantly affect the compatibility
with backscattering measurements, determining whether the new implementation of
the Goudsmit-Saunderson model in Geant4 10.2 and following patches significantly improves the
validation results over the model implemented in Geant4 10.1p03, and objectively
quantifying the capability of simulations using the Goudsmit-Saunderson model to
reproduce experimental data with respect to other physics configurations.
Contingency tables, based on the results of the Anderson-Darling test, specific
to each topic of investigation are built for this purpose over the three
energy ranges considered in this paper and analyzed by means of the tests
documented in Section \ref{sec_analysis}.

The results of the tests are reported below for Geant4 10.2p02; the same
conclusions also hold for the implementation in Geant4 10.2p01.

In each energy range, the configuration option of the Goudsmit-Saunderson model
implemented in Geant4 10.2p02 that produces the highest efficiency is taken as
a reference for comparison with other options and with the results produced with
the previous implementation in Geant4 10.1p03: the GSPWA configuration option
above 100 keV, the GSPWAERF option in the 20-100 keV range and the GSBRF option
below 20~keV.

\subsubsection{Evaluation of modeling options}

\begin{figure*}
    \centering
    \begin{subfigure}[b]{0.49\textwidth}
        \includegraphics[width=\textwidth]{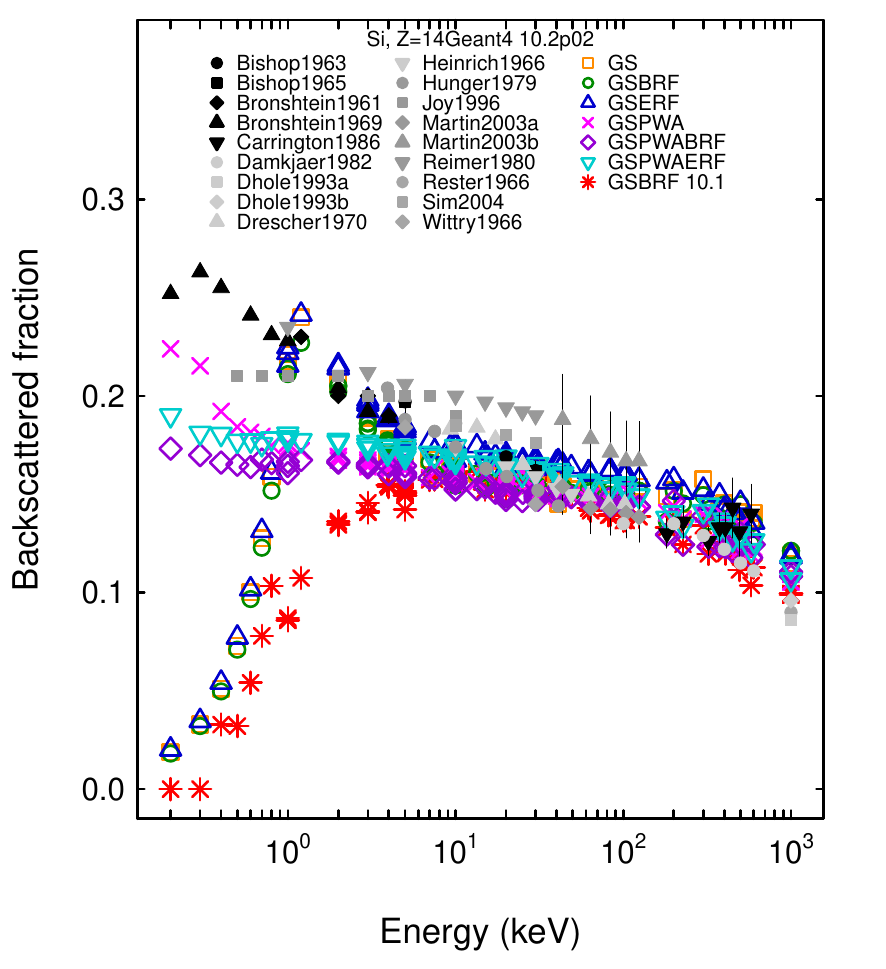}
        \label{fig_GS14}
    \end{subfigure}
    \begin{subfigure}[b]{0.49\textwidth}
        \includegraphics[width=\textwidth]{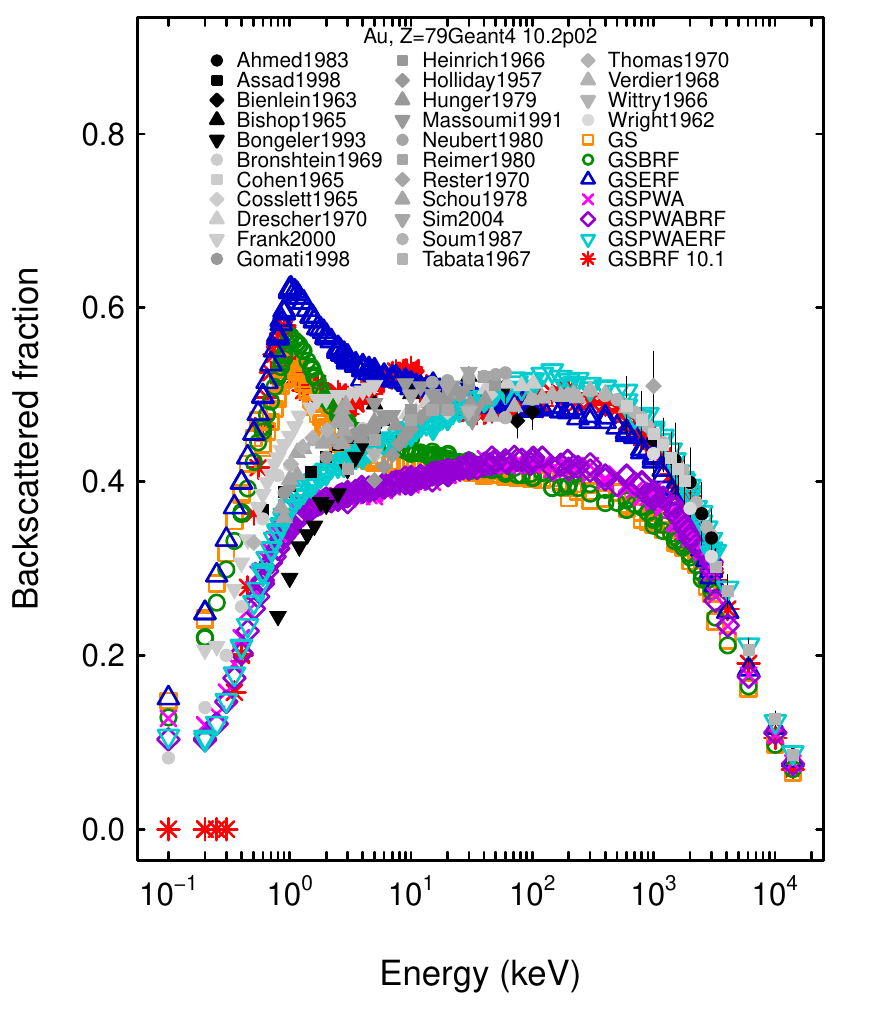}
        \label{fig_GS79}
    \end{subfigure}
    \begin{subfigure}[b]{0.49\textwidth}
        \includegraphics[width=\textwidth]{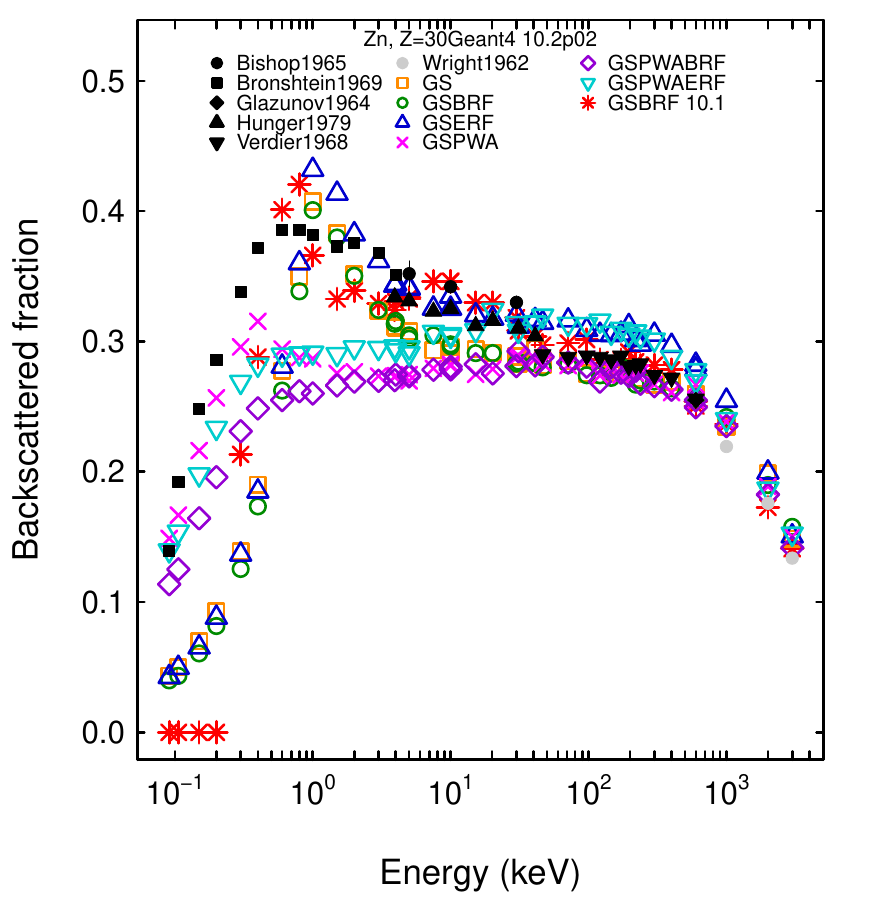}
        \label{fig_GS30}
    \end{subfigure}
    \begin{subfigure}[b]{0.49\textwidth}
        \includegraphics[width=\textwidth]{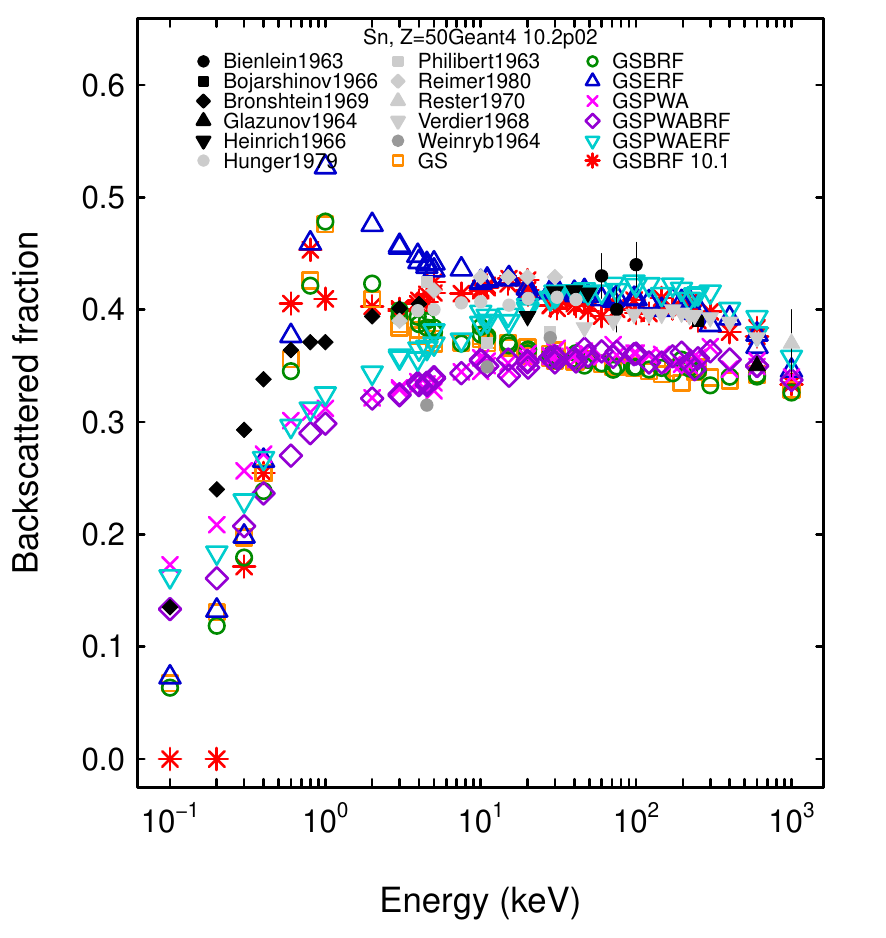}
        \label{fig_GS50}
    \end{subfigure}
\caption{Measured and simulated fraction of backscattered electrons produced 
with the Goudsmit-Saunderson multiple scattering model:
experimental data (black and grey filled markers); simulation with Geant4
10.2p02 GS (orange empty squares), GSBRF (green empty
circles), GSERF (blue empty upward triangles), GSPWA (magenta crosses), GSPWABRF
(violet empty diamonds), GSPWAERF (turquoise empty downward triangles) configurations, and
simulation with Geant4 10.1p03 GSBRF configuration (red asterisks). 
The plots concern silicon, zinc, tin and gold targets.}
\label{fig_GS}
\end{figure*}

The backscattering fraction simulated with the different options of the Goudsmit-Saunderson 
model available in Geant4 10.2p02 is illustrated in Fig.~\ref{fig_GS}
for a sample of target elements, along with experimental measurements.
The plots also report the simulation results obtained with the GSBRF
configuration in Geant4 10.1p03.

All the Goudsmit-Saunderson configurations corresponding to different screening
parameter and step limitation options available in Geant4 10.2p02 appear to
achieve similar efficiency above 100~keV, while some differences are visible at
lower energies in Table \ref{tab_eff}.
These qualitative observations are quantified through categorical data tests.
The results of testing the hypothesis of equivalent compatibility with
experiment with respect to the three reference options are documented in 
Table~\ref{tab_GS1022}.

All tests fail to reject the null hypothesis above 100~keV, while at lower
energies statistically significant differences in compatibility with experiment
are identified with respect to the modeling options that produce the largest
efficiency.

Only the GSPWAERF and GSBRF1 configuration options are not found inconsistent
with the reference configurations 
over the whole energy range covered in the validation process.
Some sensitivity to the value of the \textit{RangeFactor} parameter is observed,
as the hypothesis of equivalent compatibility with experiment with respect to
the reference option is rejected for the GSBRF option between 20 and 100 keV,
while it is not rejected for the GSBRF1 option.
It is worthwhile to note that the null hypothesis is rejected in the test
concerning 
the predefined \textit{G4EmStandardPhysicsGS} PhysicsConstructor in the energy
range between 20 and 100 keV, i.e. this simulation configuration does not
represent an optimal choice of Goudsmit-Saunderson modeling options to reproduce
experimental backscattering data at those energies.

The $\beta$-version of Geant4 recommends a lower value of the
\textit{RangeFactor} parameter (0.10 instead of 0.12) for the
Goudsmit-Saunderson multiple scattering model in the predefined
\textit{G4EmStandardPhysicsGs} PhysicsConstructor;
its effect on the compatibility of simulation with experiment obtained with the
various model options has been investigated in the same simulation context as
with Geant4 10.2p01 and is is documented in Table \ref{tab_effGS010}.
The efficiencies are qualitatively similar to those listed in Table
\ref{tab_eff}.

\tabcolsep=2pt
\begin{table*}[htbp]
  \centering
  \caption{P-values deriving from the analysis of contingency tables that compare the
compatibility with experiment obtained with the Goudsmit-Saunderson model in Geant4 10.2p02 and 10.1p03}
    \begin{tabular}{lrrrrrrrrrrrrrrr}
    \toprule
           & \multicolumn{5}{c}{Geant4 10.2p02 GSBRF $<$20 keV}  & \multicolumn{5}{c}{Geant4 10.2p02 GSPWAERF 20-100 keV} & \multicolumn{5}{c}{Geant4 10.2p02 GSPWA $>$100 keV } \\
\cmidrule(lr){2-6}
\cmidrule(lr){7-11}
\cmidrule(lr){12-16}
    \multicolumn{1}{l}{Geant4 10.1p03} & \textit{Fisher} & $\chi^2$ & \textit{Z-pooled} & \textit{Boschloo} & \textit{CSM}  & \textit{Fisher} & $\chi^2$ & \textit{Z-pooled} & \textit{Boschloo} & \textit{CSM}  & \textit{Fisher} & $\chi^2$ & \textit{Z-pooled} & \textit{Boschloo} & \textit{CSM} \\
\midrule
    GS    & 0.394 & 0.330 & 0.529 & 0.352 & 0.325 &       &       &       &       &       &          &          &          &          &          \\
    GSBRF &       &       &       &       &       & 0.689 & 0.593 & 0.682 & 0.636 & 1.000 & $<0.001$ & $<0.001$ & $<0.001$ & $<0.001$ & $<0.001$ \\
     \bottomrule
    \end{tabular}%
  \label{tab_GS1013}%
\end{table*}%
\tabcolsep=6pt

\begin{table}[htbp]
  \centering
\caption{P-values of tests of contingency tables comparing the compatibility
with experiment obtained by the GSBRF configuration with Geant4 10.1p03 and
that obtained by other physics configurations with Geant4 10.2p02, above 100
keV}
    \begin{tabular}{lrrrr}
    \toprule
    Configuration & \textit{Fisher} & \textit{Z-pooled} & \textit{Boschloo} & \textit{CSM }\\
    \midrule
    Urban       & $<0.001$ & $<0.001$ & $<0.001$ & $<0.001$ \\
    UrbanBRF    & $<0.001$ & $<0.001$ & $<0.001$ & $<0.001$ \\
    WentzelBRF  & 0.074    & 0.047    & 0.048    & 0.094    \\ 
    WentzelBRFP & 0.124    & 0.075    & 0.089    & 0.200    \\
    Coulomb     & 0.042    & 0.024    & 0.029    & 0.156    \\
    CoulombMott & 1.000    & 0.751    & 1.000    & 1.000    \\
    EmLivermore & $<0.001$ & $<0.001$ & $<0.001$ & $<0.001$ \\
    EmStd       & 0.203    & 0.126    & 0.156    & 0.471    \\
    EmOpt1      & $<0.001$ & $<0.001$ & $<0.001$ & $<0.001$ \\
    EmOpt2      & $<0.001$ & $<0.001$ & $<0.001$ & $<0.001$ \\
    EmOpt3      & 0.007    & 0.004    & 0.004    & 0.007    \\
    EmOpt4      & $<0.001$ & $<0.001$ & $<0.001$ & $<0.001$    \\
    EmWVI       & 0.074    & 0.047    & 0.048    & 0.094    \\
    EmSS        & 0.074    & 0.047    & 0.048    & 0.094    \\
    EmGS  & $<0.001$ & $<0.001$ & $<0.001$ & $<0.001$ \\
    \bottomrule
    \end{tabular}%
  \label{tab_contGS}%
\end{table}%

\tabcolsep=3pt
\begin{table*}[htbp]
  \centering
  \caption{P-values deriving from the analysis of contingency tables that compare the
compatibility with experiment obtained with the Goudsmit-Saunderson model and with other
simulation configurations in Geant4 10.2p02 below 100 keV}
    \begin{tabular}{lrrrrrrrrrr}
    \toprule
          & \multicolumn{5}{c}{Geant4 10.2p02 GSBRF $<$20 keV}  & \multicolumn{5}{c}{Geant4 10.2p02 GSPWAERF 20-100 keV}  \\
\cmidrule(lr){2-6}
\cmidrule(lr){7-11}
    \multicolumn{1}{l}{Geant4 10.1p03} & \textit{Fisher} & $\chi^2$ & \textit{Z-pooled} & \textit{Boschloo} & \textit{CSM}  & \textit{Fisher} & $\chi^2$ & \textit{Z-pooled} & \textit{Boschloo} & \textit{CSM}  \\
\midrule
    Urban       & $< 0.001$ & $< 0.001$ & $< 0.001$ & $< 0.001$ & $< 0.001$ & 0.001     & 0.001     & 0.001     & 0.001     & 0.001     \\
    UrbanBRF    & $< 0.001$ & $< 0.001$ & $< 0.001$ & $< 0.001$ & $< 0.001$ & 0.281     & 0.225     & 0.246     & 0.240     & 0.206     \\
    WentzelBRF  & $< 0.001$ & $< 0.001$ & $< 0.001$ & $< 0.001$ & $< 0.001$ & $< 0.001$ & $< 0.001$ & $< 0.001$ & $< 0.001$ & $< 0.001$ \\
    WentzelBRFP & $< 0.001$ & $< 0.001$ & $< 0.001$ & $< 0.001$ & $< 0.001$ & 0.591     & 0.502     & 0.536     & 0.536     & 0.469     \\
    Coulomb     & $< 0.001$ & $< 0.001$ & $< 0.001$ & $< 0.001$ & $< 0.001$ & 0.893     & 0.789     & 0.853     & 0.836     & 0.984     \\
    CoulombMott & $< 0.001$ &           & $< 0.001$ & $< 0.001$ & $< 0.001$ & $< 0.001$ & $< 0.001$ & $< 0.001$ & $< 0.001$ & $< 0.001$ \\
    EmLivermore & $< 0.001$ & $< 0.001$ & $< 0.001$ & $< 0.001$ & $< 0.001$ & 0.001     & 0.001     & 0.001     & 0.001     & 0.001     \\
    EmStd       & $< 0.001$ & $< 0.001$ & $< 0.001$ & $< 0.001$ & $< 0.001$ & $< 0.001$ & $< 0.001$ & $< 0.001$ & $< 0.001$ & $< 0.001$ \\
    EmOpt1      & $< 0.001$ &           & $< 0.001$ & $< 0.001$ & $< 0.001$ & $< 0.001$ &           & $< 0.001$ & $< 0.001$ & $< 0.001$ \\
    EmOpt2      & $< 0.001$ &           & $< 0.001$ & $< 0.001$ & $< 0.001$ & $< 0.001$ &           & $< 0.001$ & $< 0.001$ & $< 0.001$ \\
    EmOpt3      & $< 0.001$ & $< 0.001$ & $< 0.001$ & $< 0.001$ & $< 0.001$ & $< 0.001$ & $< 0.001$ & $< 0.001$ & $< 0.001$ & $< 0.001$ \\
    EmOpt4      & $< 0.001$ & $< 0.001$ & $< 0.001$ & $< 0.001$ & $< 0.001$ & $< 0.001$ & $< 0.001$ & $< 0.001$ & $< 0.001$ & $< 0.001$    \\
    EmWVI       & $< 0.001$ & $< 0.001$ & $< 0.001$ & $< 0.001$ & $< 0.001$ & 1.000     & 1.000     & 1.000     & 1.000     & 1.000     \\
    EmSS        & $< 0.001$ & $< 0.001$ & $< 0.001$ & $< 0.001$ & $< 0.001$ & 0.504     & 0.423     & 0.532     & 0.462     & 1.000     \\
    EmGS        & 1.000     & 0.902     & 0.949     & 1.000     & 0.861     & $< 0.001$ & $< 0.001$ & $< 0.001$ & $< 0.001$ & $< 0.001$ \\
    \bottomrule
    \end{tabular}%
  \label{tab_contGSlow}%
\end{table*}%
\tabcolsep=6pt

\tabcolsep=1pt
\begin{table*}[htbp]
  \centering
  \caption{P-values deriving from contingency tables that compare the
compatibility with experiment obtained with the most efficient predefined PhysicsConstructors in each energy range and using other
simulation configurations, with Geant4 10.2p02 }
    \begin{tabular}{lrrrrrrrrrrrrrrr}
    \toprule
          & \multicolumn{5}{c}{EmGS $<$20 keV }     & \multicolumn{5}{c}{ EmSS 20-100 keV} & \multicolumn{5}{c}{ EmStd $\ge$100 keV} \\
 \cmidrule(lr){2-6}
\cmidrule(lr){7-11}
\cmidrule(lr){12-16}
    \multicolumn{1}{l}{Configuration} & \textit{Fisher} & $\chi^2$  & \textit{Z-pooled} & \textit{Boschloo} & \textit{CSM} & \textit{Fisher} & $\chi^2$  & \textit{Z-pooled} & \textit{Boschloo} & \textit{CSM} & \textit{Fisher} & $\chi^2$  & \textit{Z-pooled} & \textit{Boschloo} & \textit{CSM} \\
   \midrule
    Urban       & $< 0.001$ & $< 0.001$ & $< 0.001$ & $< 0.001$ & $< 0.001$ & $< 0.001$ & $< 0.001$ & $< 0.001$ & $< 0.001$ & $< 0.001$ & 0.015     & 0.009     & 0.010     & 0.010     & 0.011     \\
    UrbanBRF    & $< 0.001$ & $< 0.001$ & $< 0.001$ & $< 0.001$ & $< 0.001$ & 0.060	& 0.044     & 0.048	& 0.048     & 0.047     & 0.009     & 0.005     & 0.006     & 0.006     & 0.006     \\
    GS          & 0.807     & 0.713     & 0.793     & 0.752     & 0.657     & $< 0.001$ & $< 0.001$ & $< 0.001$ & $< 0.001$ & $< 0.001$ & 0.001     & $< 0.001$ & $< 0.001$ & $< 0.001$ & $< 0.001$ \\
    GSBRF       & 1.000     & 0.902     & 0.949     & 1.000     & 0.861     & $< 0.001$ & $< 0.001$ & $< 0.001$ & $< 0.001$ & $< 0.001$ & 0.001     & $< 0.001$ & $< 0.001$ & $< 0.001$ & $< 0.001$ \\
    GSBRF1      & 0.011     & 0.008     & 0.009     & 0.009     & 0.016     & 0.285	& 0.229     & 0.250	& 0.250     & 0.571     & 0.003     & 0.002     & 0.002     & 0.002     & 0.002     \\
    GSEGSRF     & $< 0.001$ & $< 0.001$ & $< 0.001$ & $< 0.001$ & $< 0.001$ & 0.350	& 0.285     & 0.311	& 0.312     & 0.970     & 0.138     & 0.003     & 0.003     & 0.003     & 0.003     \\
    GSPWA       & $< 0.001$ & $< 0.001$ & $< 0.001$ & $< 0.001$ & $< 0.001$ & $< 0.001$ & $< 0.001$ & $< 0.001$ & $< 0.001$ & $< 0.001$ & 0.005     & 0.003     & 0.003     & 0.003     & 0.003     \\
    GSPWABRF    & $< 0.001$ & $< 0.001$ & $< 0.001$ & $< 0.001$ & $< 0.001$ & $< 0.001$ & $< 0.001$ & $< 0.001$ & $< 0.001$ & $< 0.001$ & 0.001     & $< 0.001$ & $< 0.001$ & $< 0.001$ & $< 0.001$ \\
    GSPWAEGSRF  & 0.029     & 0.021     & 0.024     & 0.024     & 0.041     & 0.504     & 0.423     & 0.532     & 0.462     & 1.000     & 0.001     & $< 0.001$ & $< 0.001$ & $< 0.001$ & $< 0.001$ \\
    WentzelBRF  & $< 0.001$ & $< 0.001$ & $< 0.001$ & $< 0.001$ & $< 0.001$ & $< 0.001$ & $< 0.001$ & $< 0.001$ & $< 0.001$ & $< 0.001$ & 0.798     & 0.607     & 0.681     & 0.653     & 1.000     \\
    WentzelBRFP & $< 0.001$ & $< 0.001$ & $< 0.001$ & $< 0.001$ & $< 0.001$ & 0.181     & 0.141     & 0.154     & 0.154     & 0.244     & 1.000     & 0.792     & 0.870     & 1.000     & 1.000     \\
    Coulomb     & $< 0.001$ & $< 0.001$ & $< 0.001$ & $< 0.001$ & $< 0.001$ & 0.350     & 0.285     & 0.311     & 0.312     & 0.970     & 0.460     & 0.324     & 0.528     & 0.395     & 0.825     \\
    CoulombMott & $< 0.001$ &           & $< 0.001$ & $< 0.001$ & $< 0.001$ & $< 0.001$ & $< 0.001$ & $< 0.001$ & $< 0.001$ & $< 0.001$ & 0.094     &           & 0.050     & 0.072     & 0.198     \\
    EmLivermore & $< 0.001$ & $< 0.001$ & $< 0.001$ & $< 0.001$ & $< 0.001$ & $< 0.001$ & $< 0.001$ & $< 0.001$ & $< 0.001$ & $< 0.001$ & 0.005     & 0.003     & 0.003     & 0.003     & 0.003     \\
    EmStd       & $< 0.001$ & $< 0.001$ & $< 0.001$ & $< 0.001$ & $< 0.001$ & $< 0.001$ & $< 0.001$ & $< 0.001$ & $< 0.001$ & $< 0.001$ &           &           &           &           &           \\
    EmOpt1      & $< 0.001$ &           & $< 0.001$ & $< 0.001$ & $< 0.001$ & $< 0.001$ &           & $< 0.001$ & $< 0.001$ & $< 0.001$ & $< 0.001$ & $< 0.001$ & $< 0.001$ & $< 0.001$ & $< 0.001$ \\
    EmOpt2      & $< 0.001$ &           & $< 0.001$ & $< 0.001$ & $< 0.001$ & $< 0.001$ &           & $< 0.001$ & $< 0.001$ & $< 0.001$ & $< 0.001$ & $< 0.001$ & $< 0.001$ & $< 0.001$ & $< 0.001$ \\
    EmOpt3      & $< 0.001$ & $< 0.001$ & $< 0.001$ & $< 0.001$ & $< 0.001$ & $< 0.001$ & $< 0.001$ & $< 0.001$ & $< 0.001$ & $< 0.001$ & 0.234     & 0.154     & 0.209     & 0.176     & 0.274     \\
    EmOpt4      & $< 0.001$ & $< 0.001$ & $< 0.001$ & $< 0.001$ & $< 0.001$ & $< 0.001$ & $< 0.001$ & $< 0.001$ & $< 0.001$ & $< 0.001$ & 0.043     & 0.025     & 0.029     & 0.029     & 0.038     \\
    EmWVI       & $< 0.001$ & $< 0.001$ & $< 0.001$ & $< 0.001$ & $< 0.001$ & 0.504     & 0.423     & 0.532     & 0.462     & 1.000     & 0.798     & 0.607     & 0.681     & 0.653     & 1.000     \\
    EmSS        & $< 0.001$ & $< 0.001$ & $< 0.001$ & $< 0.001$ & $< 0.001$ &           &           &           &           &           & 0.798     & 0.607     & 0.681     & 0.653     & 1.000     \\
    EmGS        &           &           &           &           &           & $< 0.001$ & $< 0.001$ & $< 0.001$ & $< 0.001$ & $< 0.001$ & $< 0.001$ & $< 0.001$ & $< 0.001$ & $< 0.001$ & $< 0.001$ \\
    \bottomrule
    \end{tabular}%
  \label{tab_contph}%
\end{table*}%
\tabcolsep=6pt

\subsubsection{Evaluation of the evolution of the implementation}
\label{sec_GSevolution}

\begin{figure*}
    \centering
    \begin{subfigure}[b]{0.49\textwidth}
        \includegraphics[width=\textwidth]{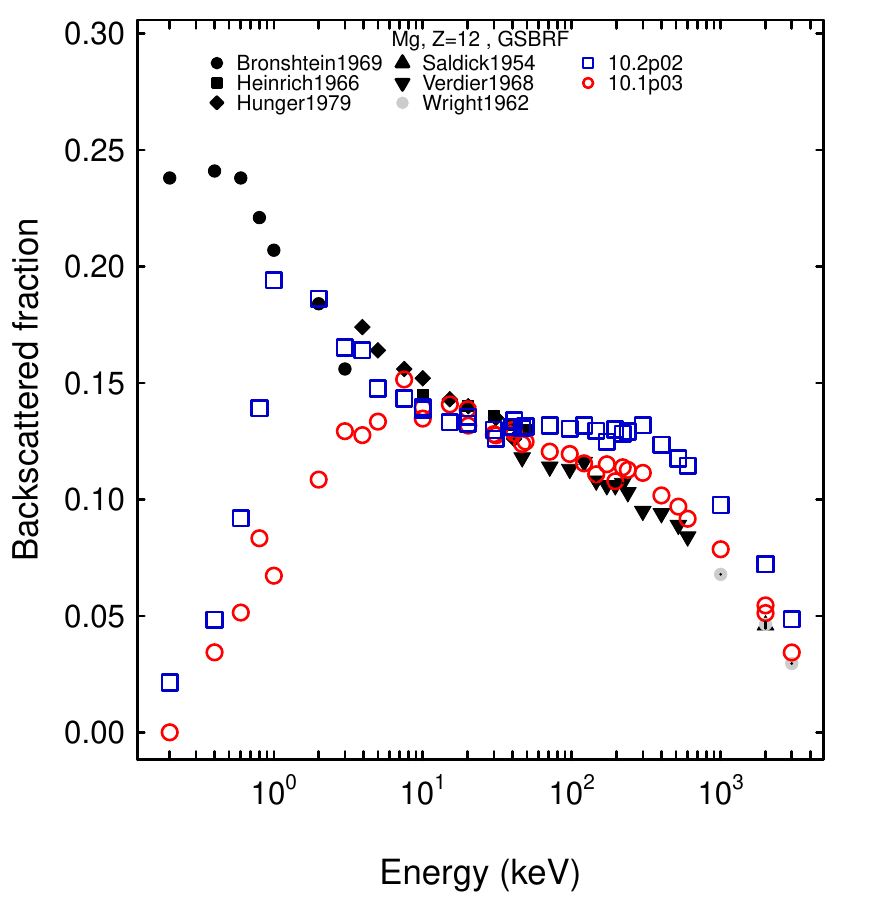}
        \label{fig_GSBRF12}
    \end{subfigure}
    \begin{subfigure}[b]{0.49\textwidth}
        \includegraphics[width=\textwidth]{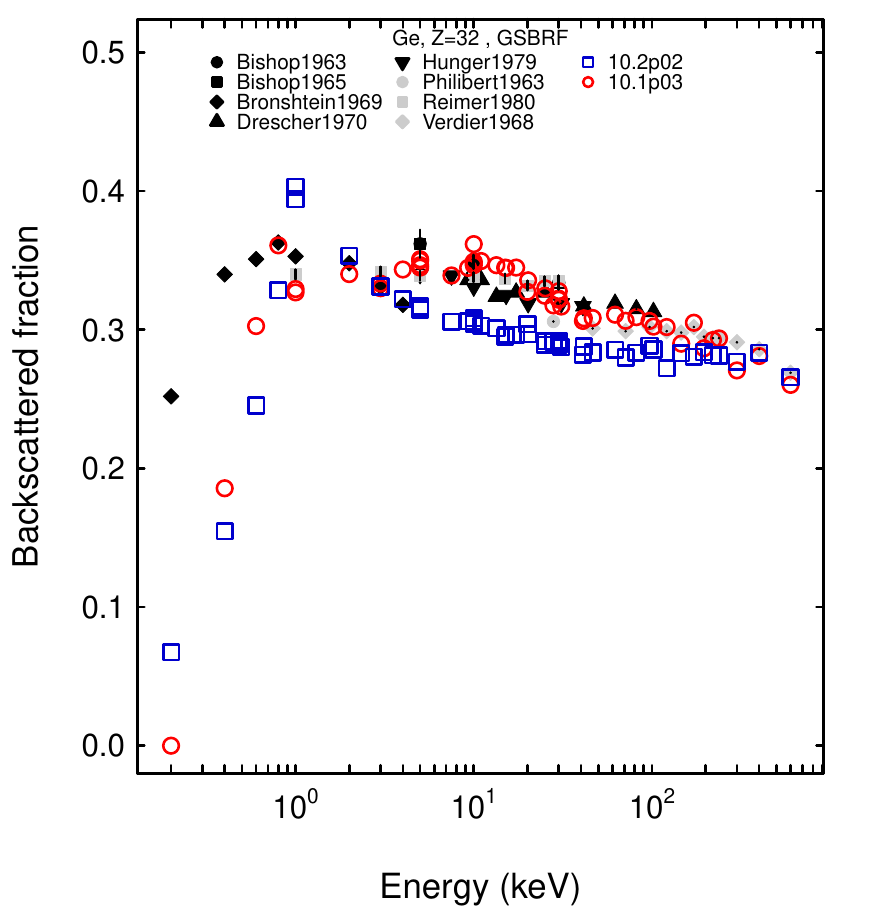}
        \label{fig_GSBRF32}
    \end{subfigure}
    \begin{subfigure}[b]{0.49\textwidth}
        \includegraphics[width=\textwidth]{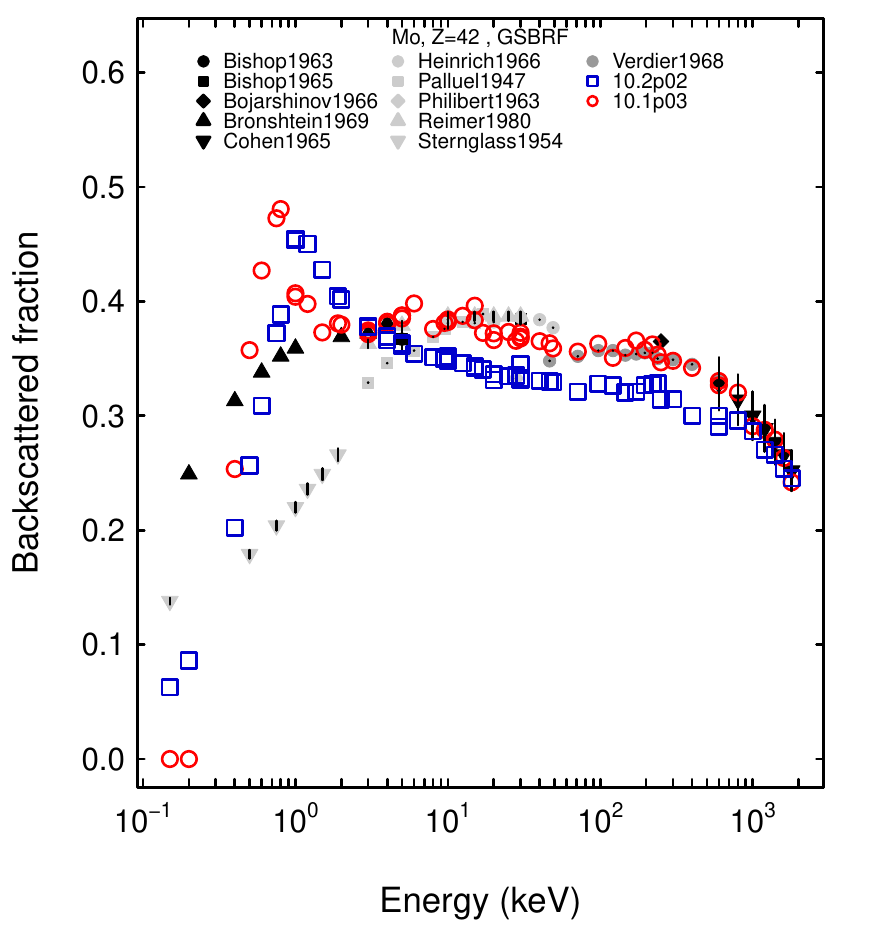}
        \label{fig_GSBRF42}
    \end{subfigure}
    \begin{subfigure}[b]{0.49\textwidth}
        \includegraphics[width=\textwidth]{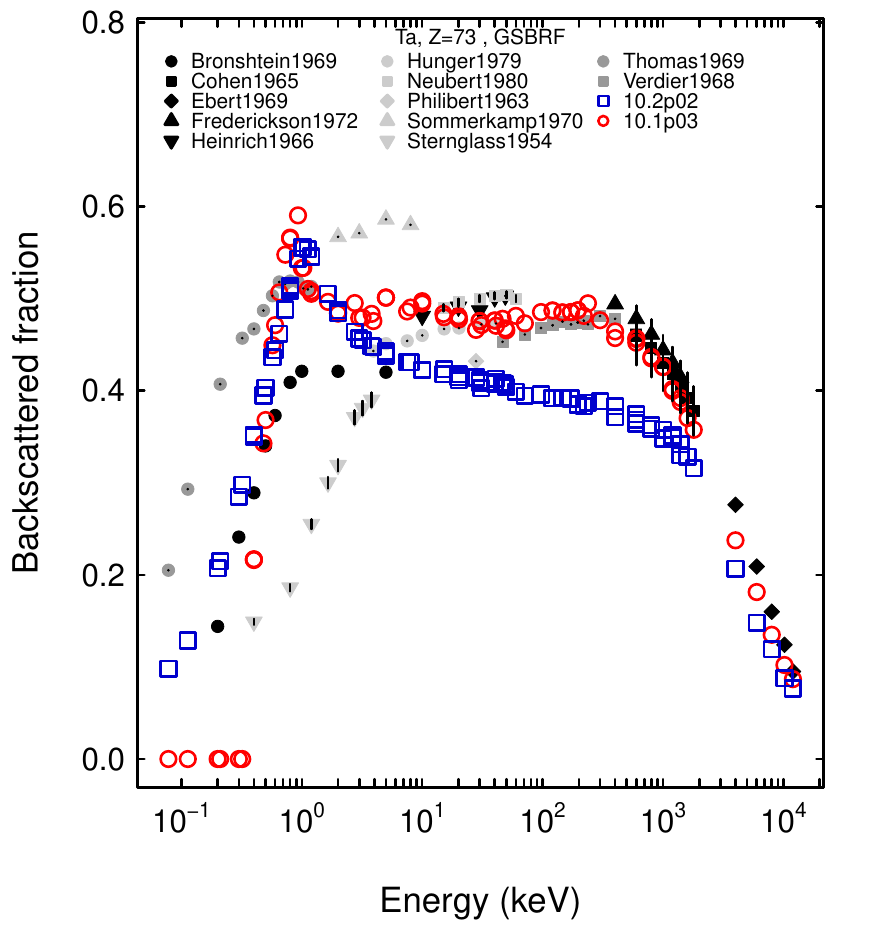}
        \label{fig_GSBRF73}
    \end{subfigure}
\caption{Measured and simulated fraction of backscattered electrons produced 
with the GSBRF configuration implementing the Goudsmit-Saunderson multiple scattering model:
experimental data (black and grey filled markers); simulation with Geant4
10.2p02  (blue empty squares) and with Geant4 10.1p03 GSBRF configuration (red empty circles). 
The plots concern magnesium, germanium, molybdenum and tantalum targets.}
\label{fig_GSBRF}
\end{figure*}

The implementation of the Goudsmit-Saunderson multiple scattering model
does not appear do have consistently evolved towards better compatibility 
with experiment from Geant4 10.1p03 to 10.2p02.
Fig.~\ref{fig_GSBRF} illustrates some examples of the evolution associated 
with the GSBRF simulation configuration.

A substantial drop in efficiency above 100~keV is observed in Table
\ref{tab_eff} with the Goudsmit-Saunderson multiple scattering implementation in
Geant4 10.2 with respect to the value achieved with the implementation in 
Geant4 10.1p03, irrespective of the options selected for the calculation of the
screening parameter and for the step limitation algorithm.
The deterioration of compatibility with experiment is confirmed in the results
obtained with the 10.2p01 and 10.2p02 versions.
This decrease does not appear to be due to the larger value of the
\textit{RangeFactor} parameter recommended for the new implementation, as
comparable results are obtained with the recommended value of 0.12 and with the
value of 0.01 set in the GSBRF1 configuration, which is the same as in the
simulation with the GSBRF configuration in the Geant4 10.1p03 environment.
Since the efficiencies associated with simulation configurations involving other
multiple scattering models (Urban and WentzelVI) remain statistically equivalent
above 100~keV over the Geant4 versions considered in Table \ref{tab_eff}, it is
unlikely that the degradation of compatibility with experiment could originate
from evolutions in Geant4 kernel code other than the implementation of
Goudsmit-Saunderson multiple scattering.

The evolution of the capability to reproduce backscattering measurements is
quantified through the test of contingency tables, which compare the
compatibility with experiment achieved in each energy range by the
Goudsmit-Saunderson configuration options associated with the highest efficiency
in the Geant4 10.2p02 and 10.1p03 environment, respectively.
Table \ref{tab_GS1013} reports the results of this test.
The null hypothesis of equivalent compatibility with experiment  is rejected 
above 100 keV, while it is not rejected at lower energies.

No substantial change regarding the comparison with the results deriving from
Geant4 10.1.p03 is observed in the analysis of contingency tables produced with
the lower \textit{RangeFactor} value of 0.10.


From this analysis one can infer that the new implementation of the
Goudsmit-Saunderson multiple scattering model, first released in Geant4 10.2, is
equivalent to the previous one at reproducing experimental backscattering data
below 100 keV, while it has negatively improved
the compatibility of simulation with experiment above 100~keV.

Compatibility with experiment that is statistically equivalent to that obtained
with the GSBRF configuration in the Geant4 10.1p03 environment can be achieved
with configurations other than the Goudsmit-Saunderson options with Geant4
10.2p02 above 100~keV.
The results of this analysis are summarized in Table \ref{tab_contGS}, which
reports the outcome of the tests of contingency tables involving the GSBRF
configuration of Geant4 10.1p03 and other physics configurations pertinent to
Geant4 10.2p02: the null hypothesis is not rejected for configurations including
single scattering models and the WentzelVI model, either in user-defined or
predefined Phy\-sics\-Con\-structors, and for the EmStd configuration.



\subsubsection{Evaluation with respect to other electron scattering models}

This analysis is focused on the lower energy end, where simulation
configurations encompassing the Goudsmit-Saunderson multiple scattering model
are associated with relatively high efficiencies in the Geant4 10.2p02
environment.

Similarly to the previously documented evaluations, the GSBRF and GSPWAERF
configurations, which achieve the largest efficiencies below 20 keV and in the
20-100 keV range, respectively, are considered as references in contingency
tables, which compare their compatibility with experiment with that obtained
with other simulation configurations in the context of Geant4 10.2p02.
The results of these tests are summarized in Table \ref{tab_contGSlow}.
Below 20 keV, the hypothesis of equivalent compatibility with experiment is rejected for
all configurations with respect to the GSBRF configurations.

In the intermediate energy range the hypothesis of equivalent compatibility with
experiment is not rejected for the UrbanBRF and WentzelBRF configurations, nor
when using single scattering with the \textit{G4eCoulombScatteringModel} instead
of multiple scattering for electron transport.
Consistent results are obtained regarding the EmSS and EmWVI predefined 
PhysicsConstructors, which use the WentzelVI and single Coulomb scattering models.

This analysis indicates that the Goudsmit-Saunderson implementation of Geant4
10.2p02 is significantly better than the other configurations subject to test at
simulating electron backscattering below 20~keV, while it is statistically
equivalent to other multiple and single scattering configurations available in
the Geant4 10.2p02 environment in the 20-100~keV energy range.
It should be noted, however, that the efficiencies at lower energies are
substantially lower than those achieved above 100 keV by the most efficient
configurations.

The lower \textit{RangeFactor} value foreseen for Geant4 10.3-beta,
mentioned in the previous subsection \ref{sec_GSevolution}, does not 
substantially change the outcome of the comparisons with other electron 
scattering models.

\subsection{Single Scattering Model Based on the Mott Cross Section}
\label{sec_mott}

No backscattered electrons are observed in simulations using the single
scattering model based on the Mott cross section included in Geant4 10.2.
The cause of this anomaly
was corrected in the version released in Geant4 10.2p01.

The efficiency of simulations involving the CoulombMott configuration is
documented in Table \ref{tab_eff}: it is low below 100~keV, consistent with
the documented model specifications, while it is the highest achieved with
Geant4 10.2p01 by any of the simulation configurations subject to evaluation 
above 100~keV.
Statistical tests do not have sufficient discriminant power to appraise
differences in compatibility with experiment between 100 keV and the nominal
lower applicability limit of 200 keV due to the scarce amount of
experimental data available in this energy range.
Although the Geant4 documentation says that the
\textit{G4eSingleCoulombScatteringModel} model is applicable to scattering
between electrons and medium-light nuclei, no significant difference related to
the target atomic number is observed regarding the capability to reproduce
the experimental fraction of backscattered electrons.

\begin{figure}
    \centering
    \begin{subfigure}[b]{0.49\textwidth}
        \includegraphics[width=\textwidth]{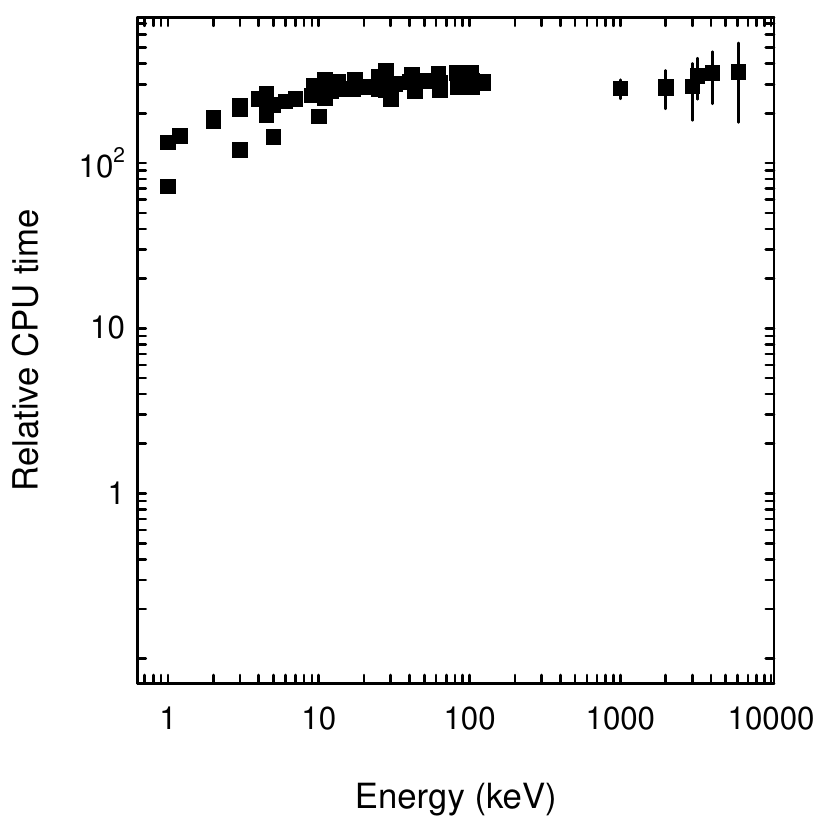}
        \label{fig_timeCoulombRatio4}
    \end{subfigure}

    \begin{subfigure}[b]{0.49\textwidth}
        \includegraphics[width=\textwidth]{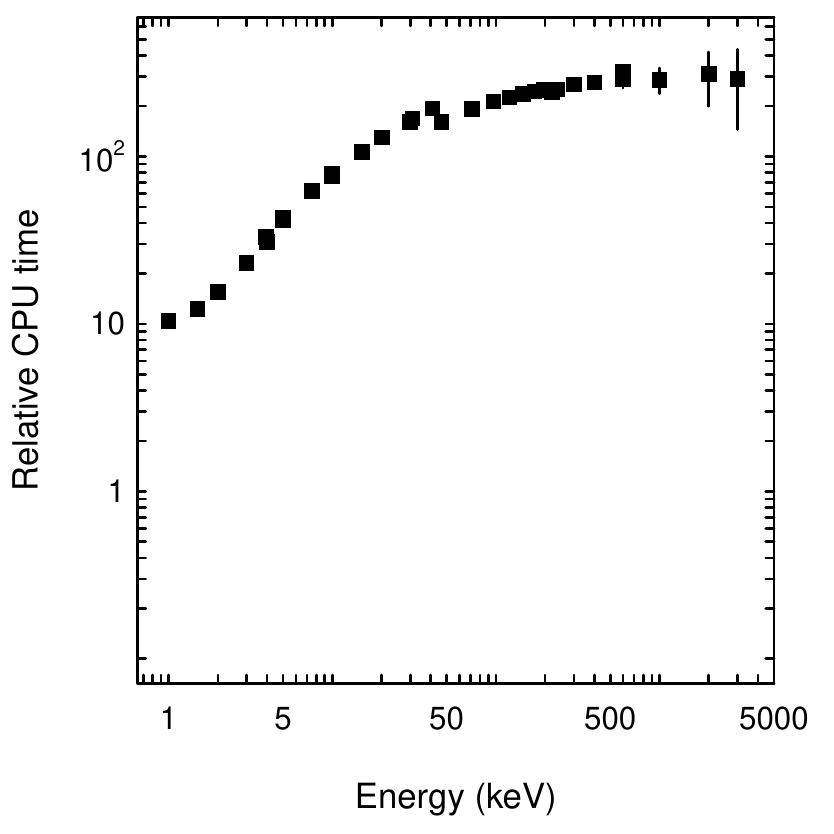}
        \label{fig_timeCoulombRatio30}
    \end{subfigure}
\caption{
Ratio of average CPU
time per event for simulations involving G4eSingleCoulombScatteringModel
and G4eCoulombScatteringModel in Geant4 10.2p01.
The test cases concern beryllium (top) and zinc (bottom) targets.
Statistical uncertainties that are smaller than the size of the markers are
not visible. As explained in the text, the plot has a qualitative character, since
the characteristics of the production of the backscattering validation test are
not suitable for rigorous computational performance estimates.}
\label{fig_timeCoulombRatio}
\end{figure}

\begin{figure}
    \centering
    \begin{subfigure}[b]{0.49\textwidth}
        \includegraphics[width=\textwidth]{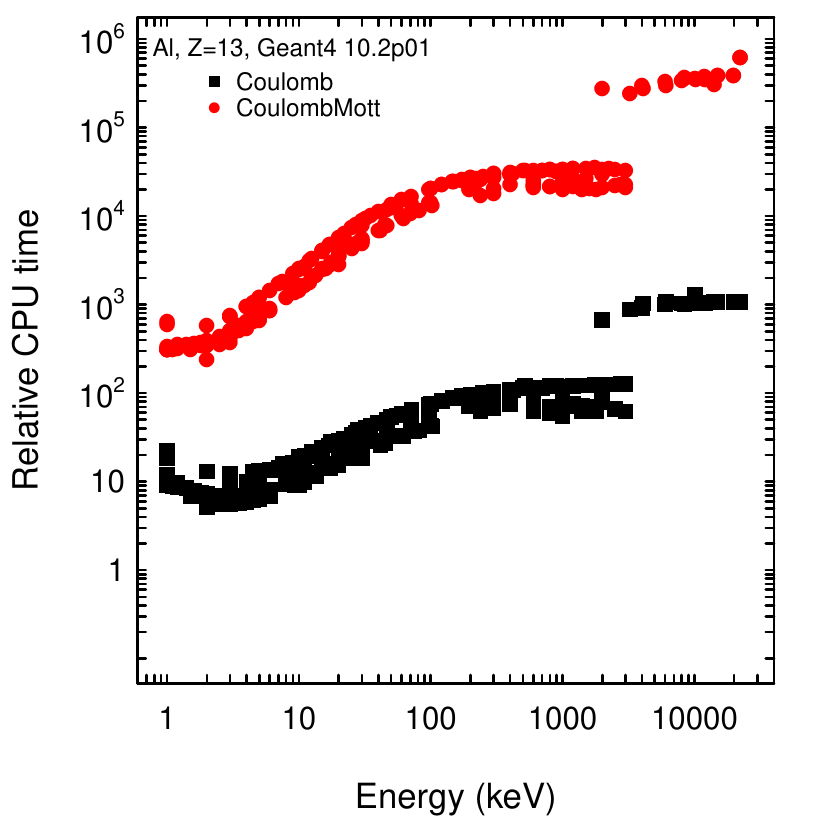}
        \label{fig_timeCoulomb13}
    \end{subfigure}

    \begin{subfigure}[b]{0.49\textwidth}
        \includegraphics[width=\textwidth]{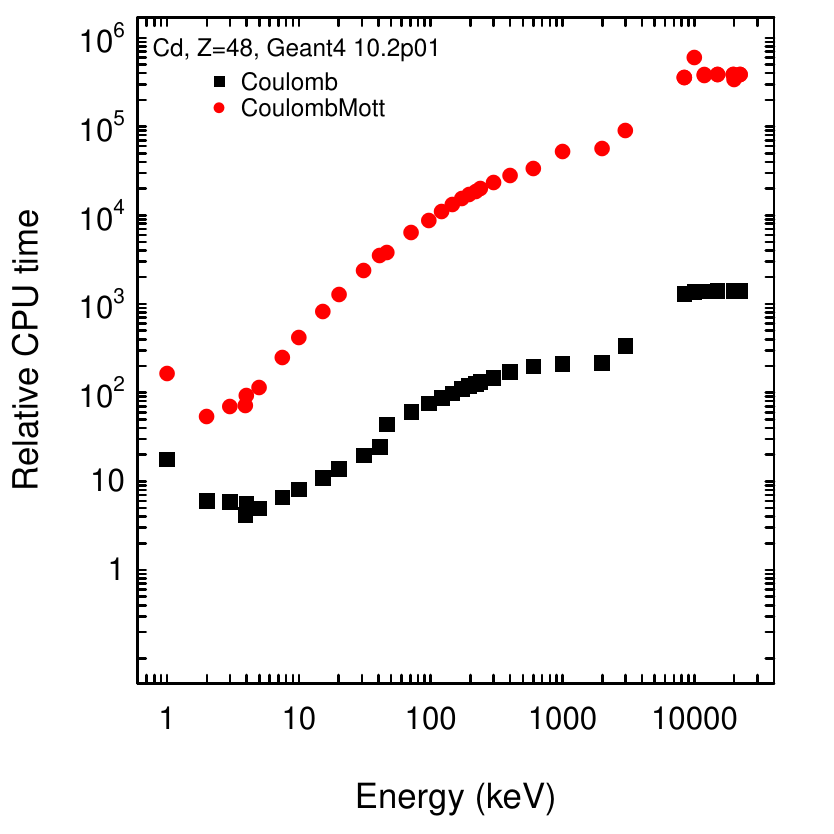}
        \label{fig_timeCoulomb48}
    \end{subfigure}
\caption{
Ratio of average CPU
time per event for configurations including G4eCoulombScatteringModel (Coulomb, black
squares) and G4eSingleCoulombScatteringModel (CoulombMott, red circles), with respect
to the configuration including G4EmStandardPhysics in Geant4 10.2p01.
Statistical uncertainties that are smaller than the size of the markers  are
not visible. As explained in the text, the plot has a qualitative character, since
the characteristics of the production of the backscattering validation test are
not suitable for rigorous computational performance estimates.}
\label{fig_timeCoulomb}
\end{figure}

The modifications implemented in Geant4 10.2p01 to address physical correctness
have severely affected the computational performance of the simulation.
As illustrated in Fig.~\ref{fig_timeCoulombRatio}, simulations involving the
CoulombMott configuration in the Geant4 10.2p01 environment are approximately
two orders of magnitude slower than those involving the previously existing
Geant4 single scattering model \textit{G4eCoulombScatteringModel}, which in turn
imposes a significant penalty with respect to multiple scattering models.
Fig. \ref{fig_timeCoulomb} shows some examples of the computational performance 
associated with the Coulomb and CoulombMott configurations, with respect to that
of simulations using the \textit{G4EmStandardPhysics} PhysicsConstructor, which
involves multiple scattering modeling.
The computational performance of simulations using the
\textit{G4eCoulombScatteringModel} is de facto prohibitively slow for practical
use in experimental scenarios similar to the simple setup modeled in this
validation test.

\begin{figure} 
\centerline{\includegraphics[angle=0,width=8.5cm]{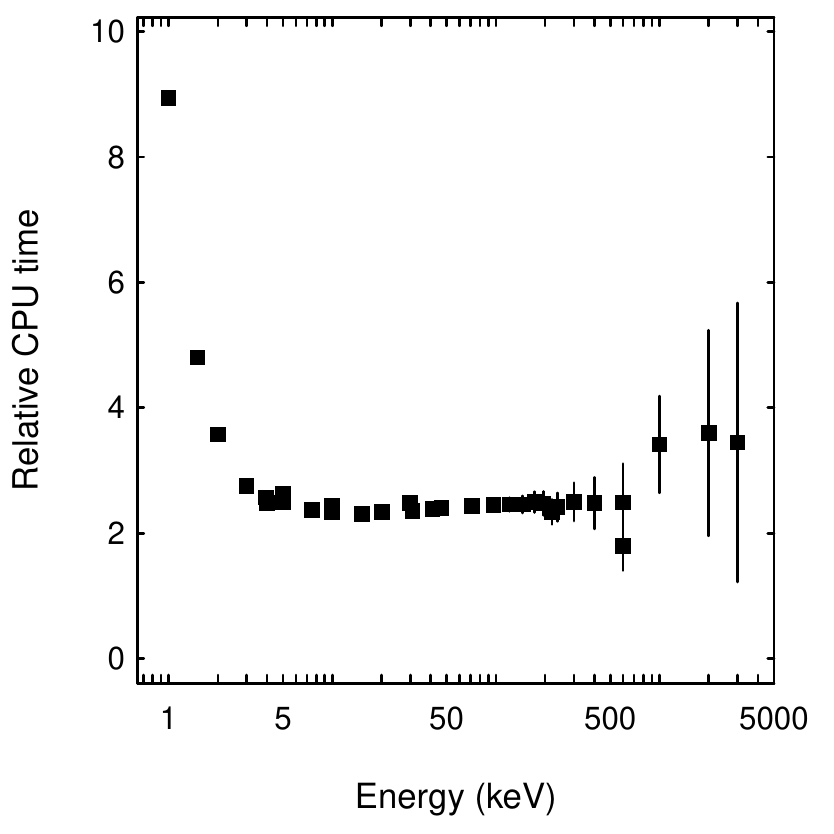}}
\caption{Factor of improvement of the computational performance of
the CoulombMott simulation configuration for a zinc target, as a function of the
primary electron energy; the plot represents the ratio of CPU time consumption
with Geant4 10.2p01 and with an improved version of the
\textit{G4eSingleCoulombScattering} model in Geant4 10.2p02. }
\label{fig_CoulombMiFactor30}
\end{figure}

Modifications of \textit{G4eCoulombScatteringModel} im\-ple\-men\-ted
in Geant4 10.2p02 
have improved the simulation speed by approximately a factor 2.5 without
affecting its compatibility with experimental data; an example of the
computational performance improvement achieved with Geant4 10.2p02 with respect
to Geant4 10.2p01 is shown in Fig.~\ref{fig_CoulombMiFactor30}.
Nevertheless, this improvement has limited impact on the practical usability of
this model in experimental applications, as such a speed increase
would not substantially change the situation depicted in Fig.~\ref{fig_timeCoulombRatio}.


\subsection{Predefined Electromagnetic PhysicsConstructors}

\begin{figure*}
    \centering
    \begin{subfigure}[b]{0.49\textwidth}
        \includegraphics[width=\textwidth]{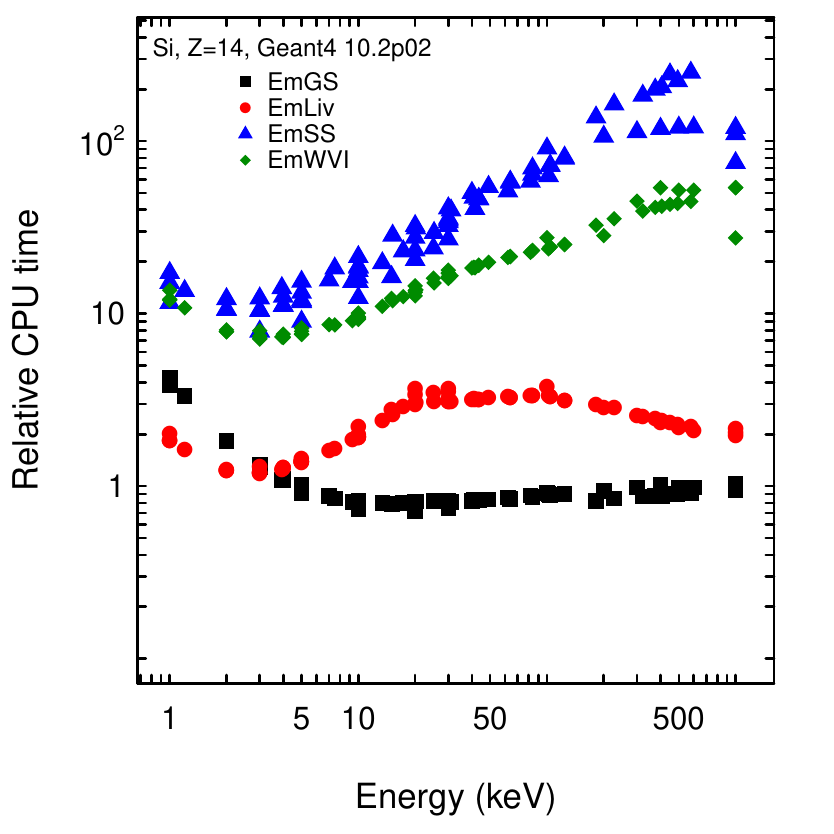}
        \label{fig_time14}
    \end{subfigure}
    \begin{subfigure}[b]{0.49\textwidth}
        \includegraphics[width=\textwidth]{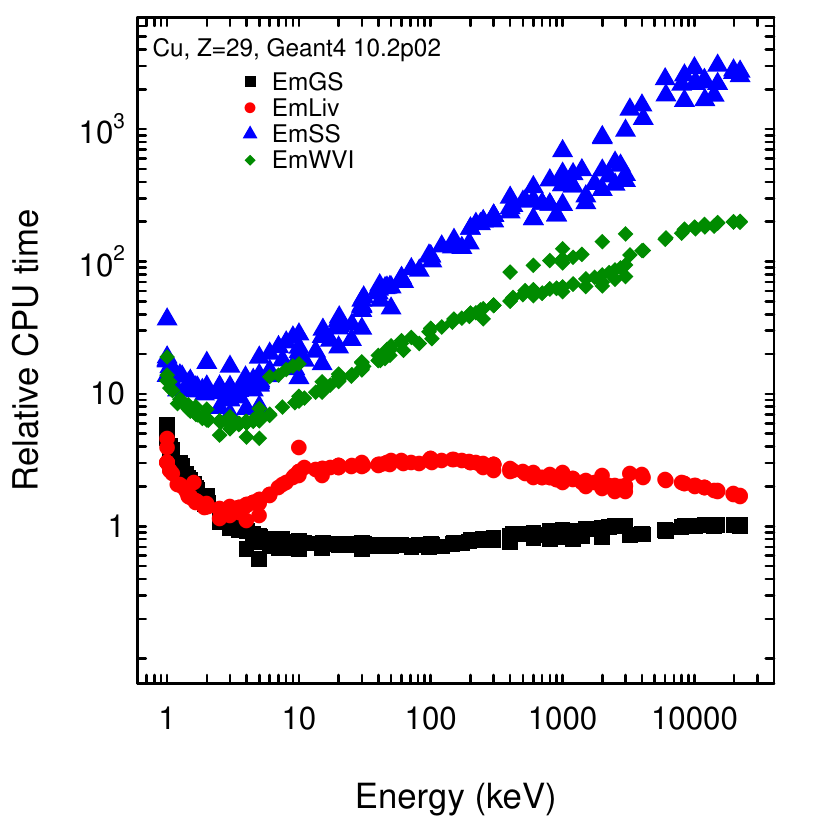}
        \label{fig_time29}
    \end{subfigure}

    \begin{subfigure}[b]{0.49\textwidth}
        \includegraphics[width=\textwidth]{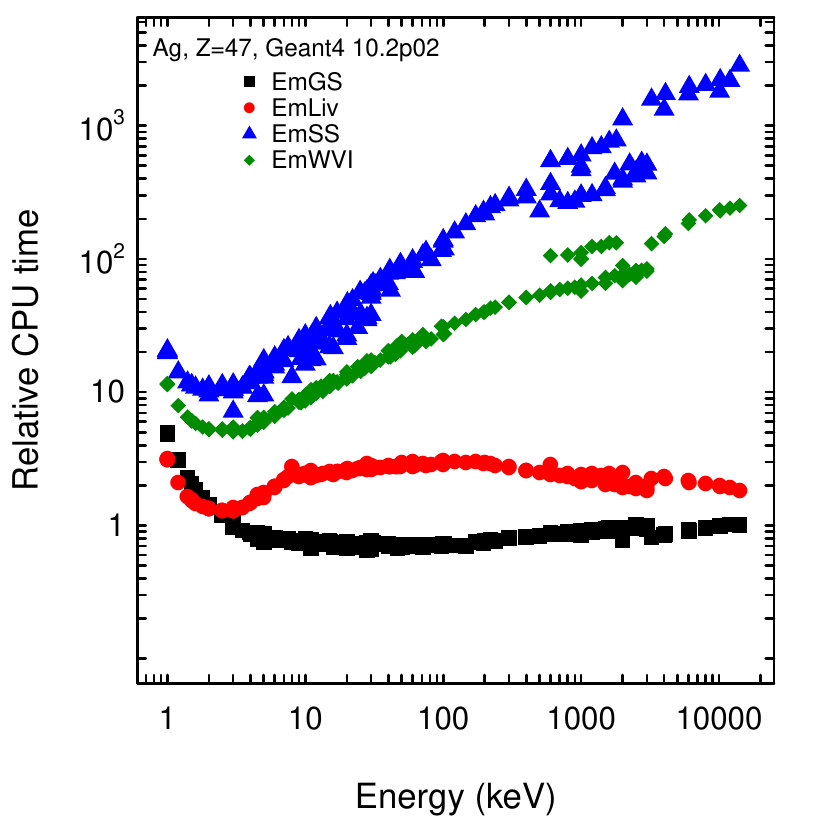}
        \label{fig_time47}
    \end{subfigure}
    \begin{subfigure}[b]{0.49\textwidth}
        \includegraphics[width=\textwidth]{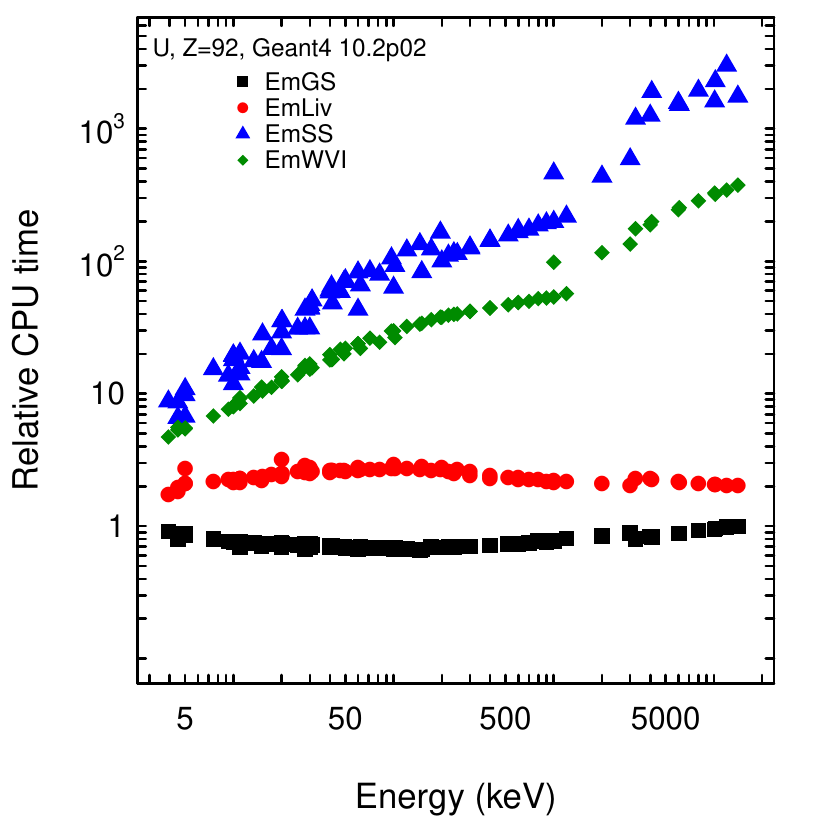}
        \label{fig_time92}
    \end{subfigure}
\caption{Computational performance of simulation configurations including
predefined electromagnetic PhysicsConstructors, in test cases involving silicon,
copper, silver and uranium targets: the plots report the ratio of average CPU
time per event for configurations including , with respect
to the configuration including G4EmStandardPhysics.
The simulation were performed with Geant4 10.2p02.
Statistical uncertainties are smaller than the size of the markers, if they are
not visible. As explained in the text, the plot has a qualitative character, since
the characteristics of the production of the backscattering validation test are
not suitable for rigorous computational performance estimates.}
\label{fig_time}
\end{figure*}

\begin{figure*}
    \centering
    \begin{subfigure}[b]{0.49\textwidth}
        \includegraphics[width=\textwidth]{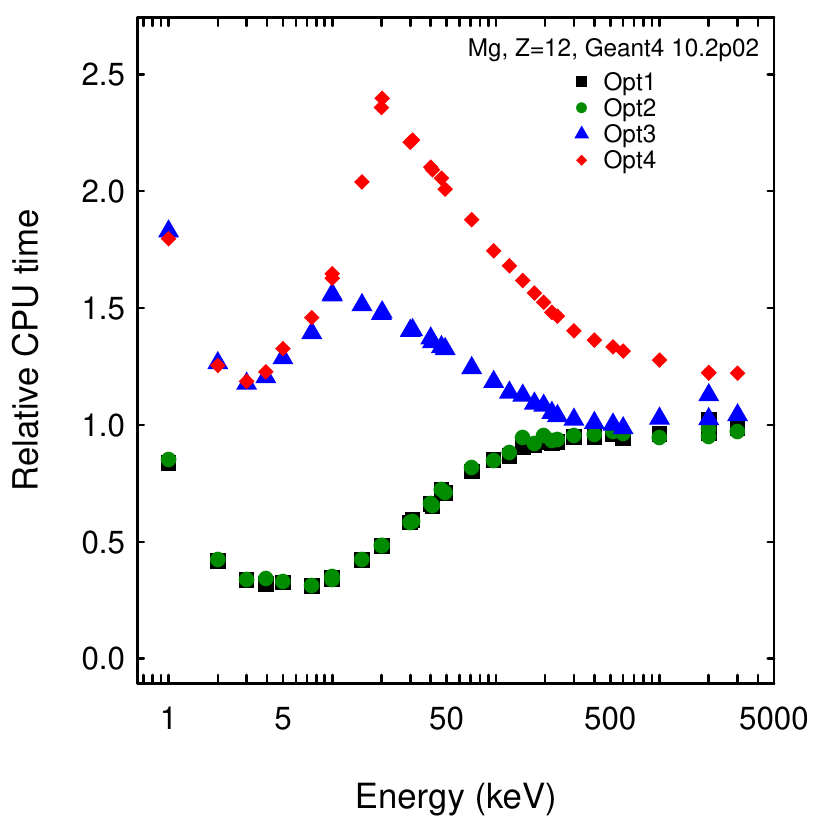}
        \label{fig_timeStd12}
    \end{subfigure}
    \begin{subfigure}[b]{0.49\textwidth}
        \includegraphics[width=\textwidth]{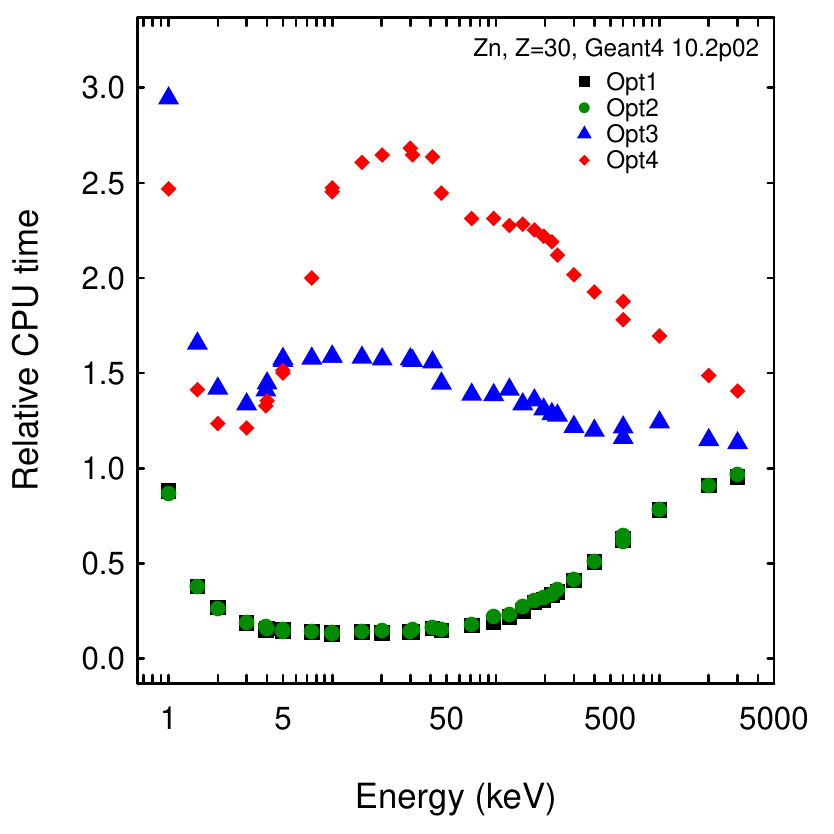}
        \label{fig_timeStd30}
    \end{subfigure}

    \begin{subfigure}[b]{0.49\textwidth}
        \includegraphics[width=\textwidth]{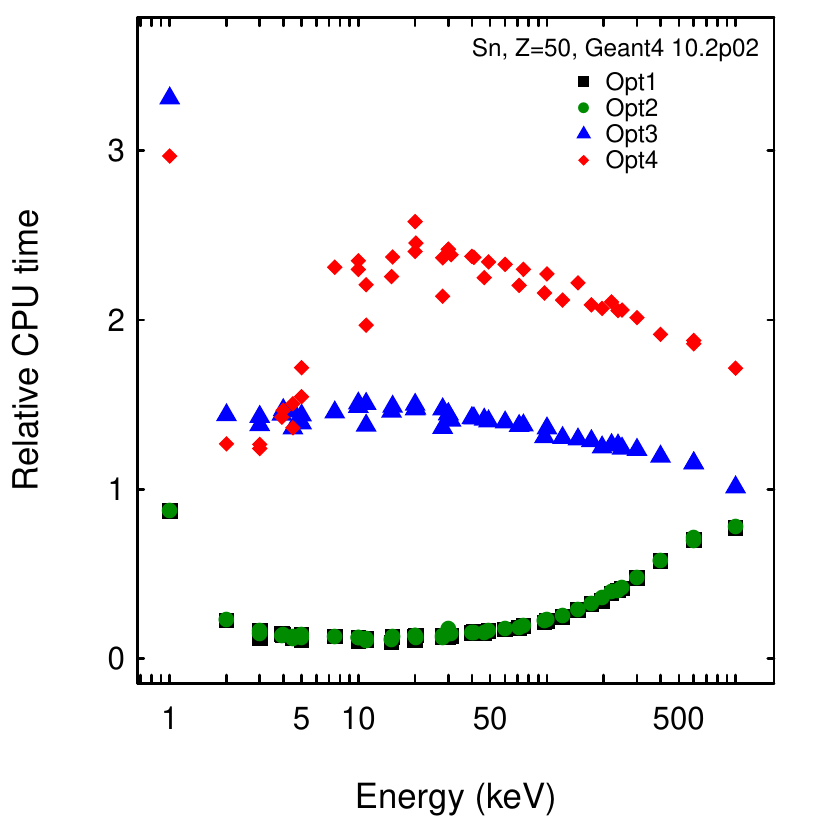}
        \label{fig_timeStd50}
    \end{subfigure}
    \begin{subfigure}[b]{0.49\textwidth}
        \includegraphics[width=\textwidth]{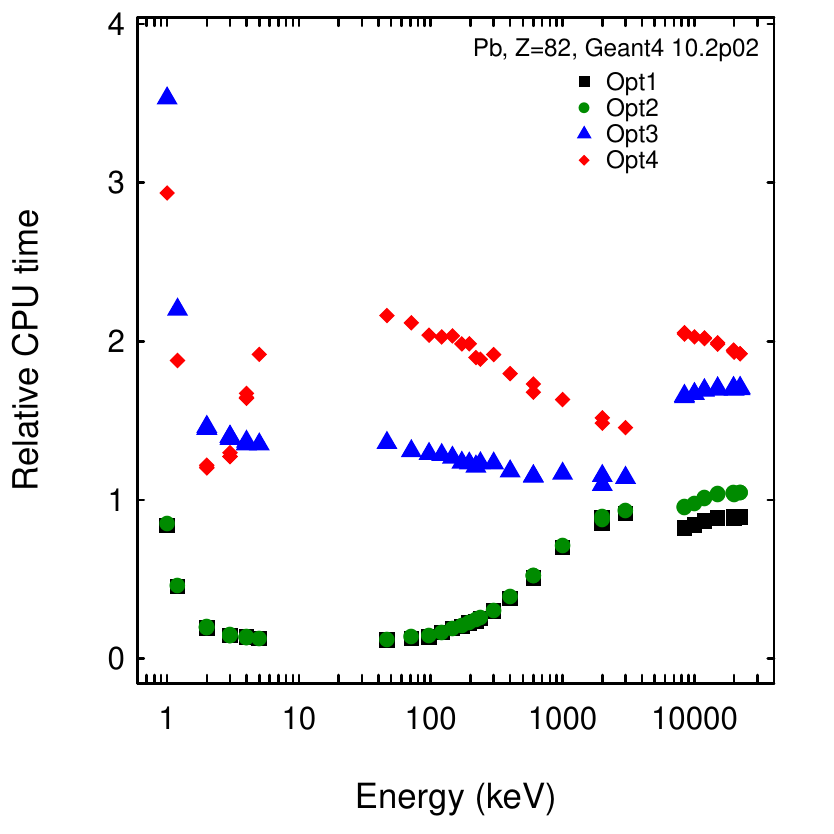}
        \label{fig_timeStd82}
    \end{subfigure}
\caption{Computational performance of simulation configurations including
predefined electromagnetic PhysicsConstructors, in test cases involving magnesium,
zinc, tin and lead targets: the plots report the ratio of average CPU
time per event for configurations including G4EmStandardPhysics\_option1 (EmOpt1, black
squares), G4EmStandardPhysics\_option2 (EmOpt2, green circles), G4EmStandardPhysics\_option3 (EmOpt3,
blue triangles) and G4EmStandardPhysics\_option4 (EmOpt4, red diamonds), with respect
to the configuration including G4EmStandardPhysics.
The simulation were performed with Geant4 10.2p02.
Statistical uncertainties that are smaller than the size of the markers are
not visible. As explained in the text, the plot has a qualitative character, since
the characteristics of the production of the backscattering validation test are
not suitable for rigorous computational performance estimates.}
\label{fig_timeStd}
\end{figure*}

The backscattering fraction simulated with different predefined electromagnetic
PhysicsConstructors released in Geant4 10.2p01 is illustrated in
Figs.~\ref{fig_phGS} and \ref{fig_phOpt} for a set of target elements,
along with experimental measurements.
Complementary information regarding the associated computational performance
can be found in Figs. \ref{fig_time}  and \ref{fig_timeStd}.
Both figures show the ratio of CPU time with respect to simulations with the EmStd 
configuration; 
Fig. \ref{fig_time} concerns the EmGS, EmLivermore, EmSS and EmWVI configurations,
while Fig. \ref{fig_timeStd} concerns the four EmOpt1-EmOpt4 options available
as variants of the \textit{G4EmStandardPhysics} predefined PhysicsConstructor.

\subsubsection{Compatibility with experiment}

\begin{figure*}
    \centering
    \begin{subfigure}[b]{0.49\textwidth}
        \includegraphics[width=\textwidth]{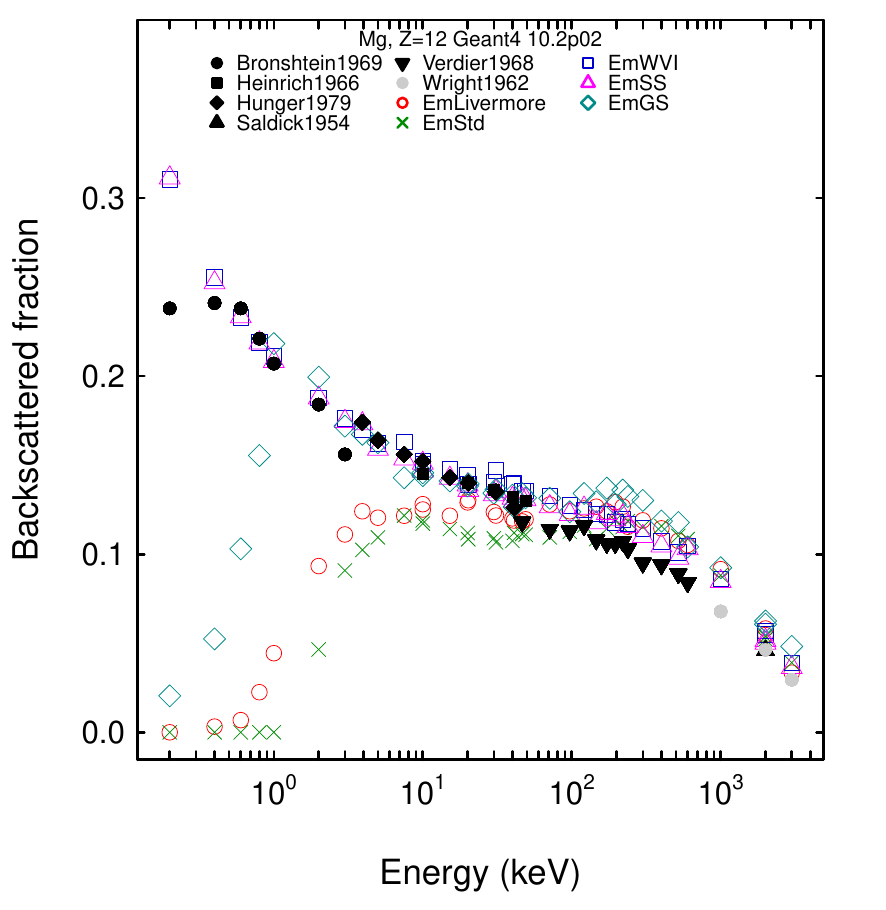}
        \label{fig_physListGS_v1022_12}
    \end{subfigure}
    \begin{subfigure}[b]{0.49\textwidth}
        \includegraphics[width=\textwidth]{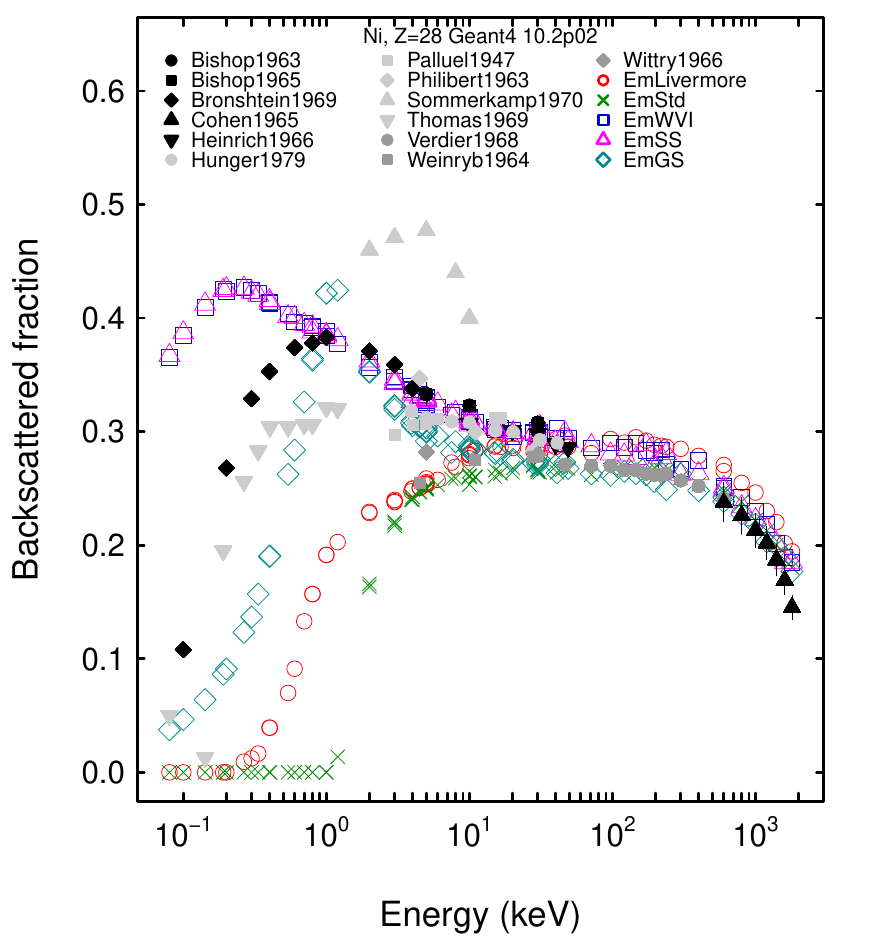}
        \label{fig_physListGS_v1022_12}
    \end{subfigure}
    \begin{subfigure}[b]{0.49\textwidth}
        \includegraphics[width=\textwidth]{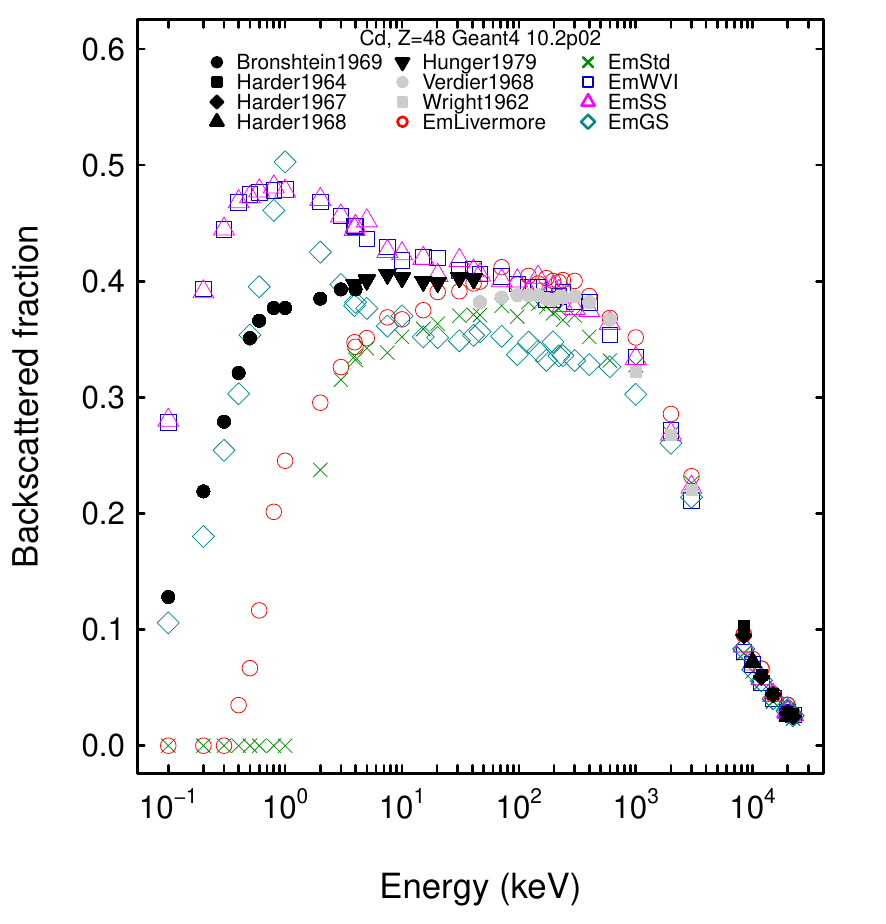}
        \label{fig_physListGS_v1022_48}
    \end{subfigure}
    \begin{subfigure}[b]{0.49\textwidth}
        \includegraphics[width=\textwidth]{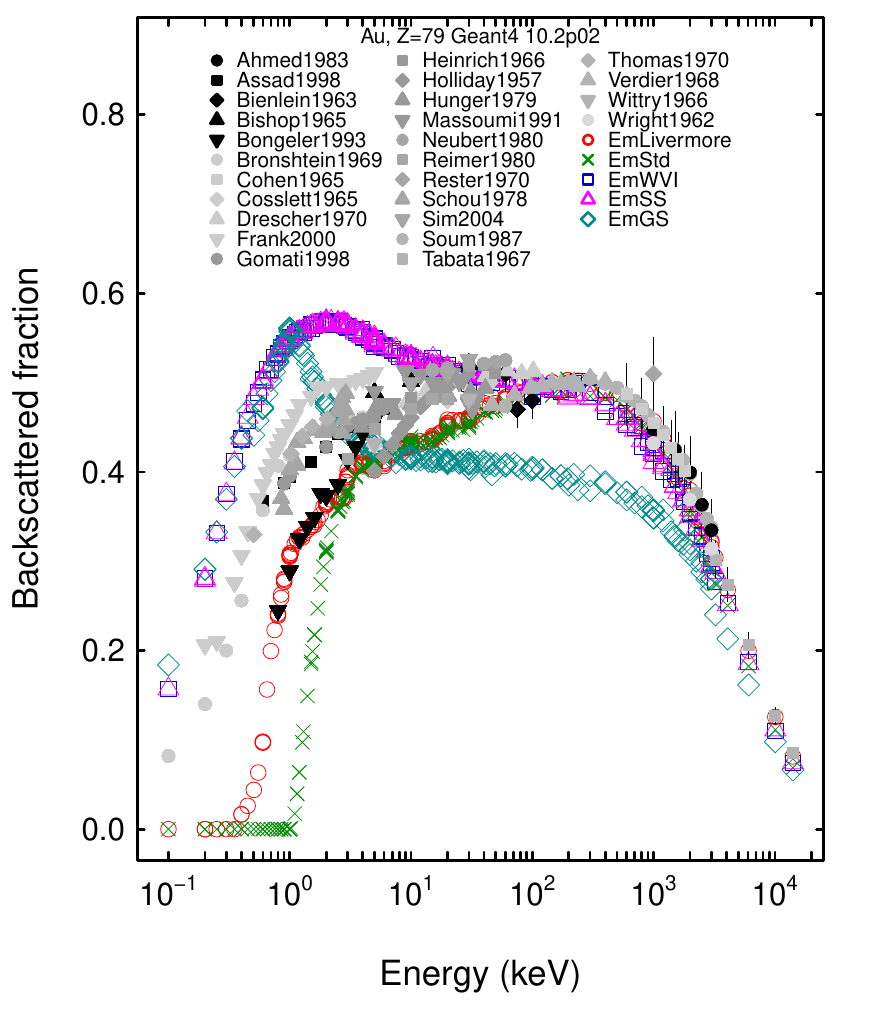}
        \label{fig_physListGS_v1022_79}
    \end{subfigure}
\caption{Measured and simulated fraction of backscattered electrons produced 
with different predefined PhysicsConstructors:
experimental data (black and grey filled markers); simulation with Geant4 10.2p02 
\textit{G4EmStandardPhysics} (EmStd, green crosses), 
\textit{G4EmStandardPhysicsGS} (EmGS turquoise empty diamonds), 
\textit{G4EmStandardPhysicsSS} (blue empty squares), 
\textit{G4EmStandardPhysicsWVI} (magenta empty triangles), 
\textit{G4EmLivermorePhysics}  (red empty circles). 
The plots concern magnesium, nickel, cadmium and gold targets.}
\label{fig_phGS}
\end{figure*}

\begin{figure*}
    \centering
    \begin{subfigure}[b]{0.49\textwidth}
        \includegraphics[width=\textwidth]{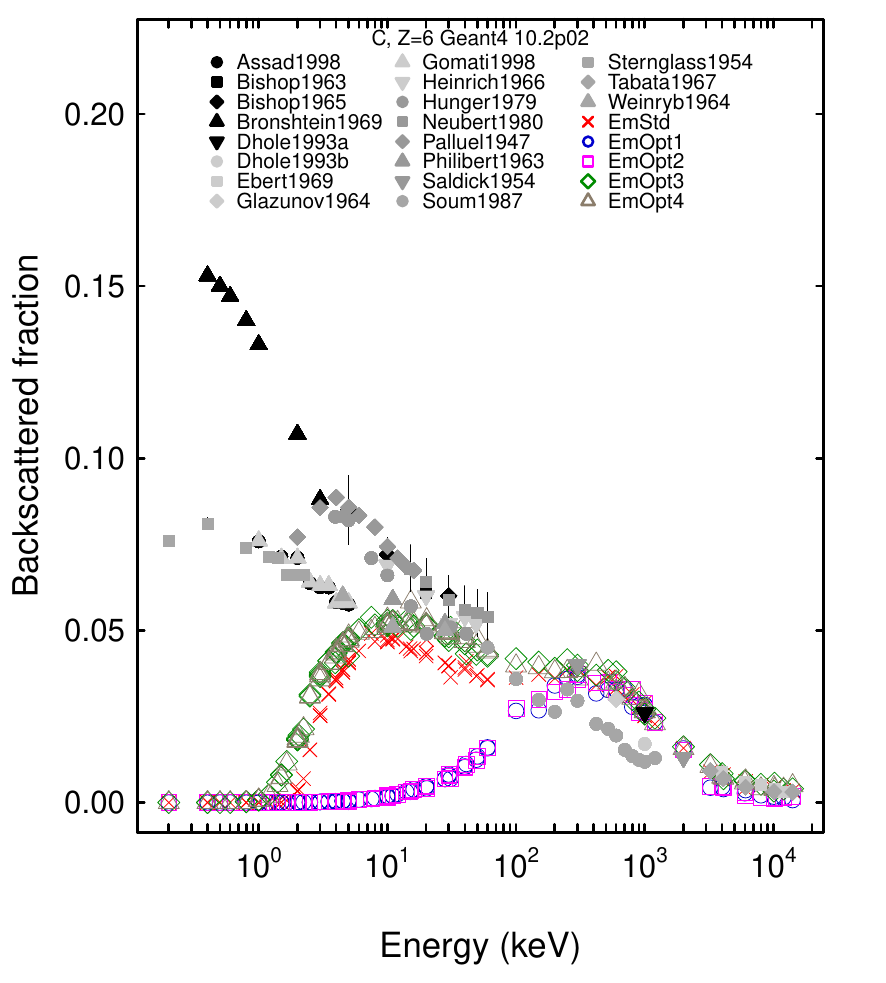}
        \label{fig_physListStd_v1022_6}
    \end{subfigure}
    \begin{subfigure}[b]{0.49\textwidth}
        \includegraphics[width=\textwidth]{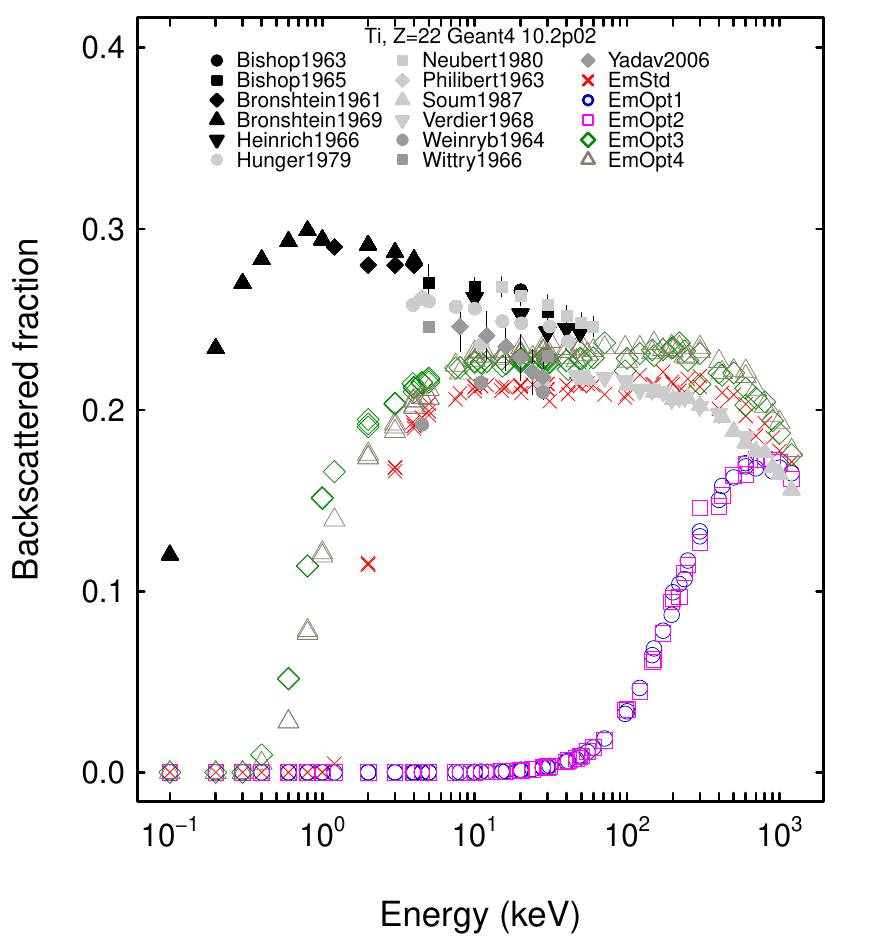}
        \label{fig_physListStd_v1022_22}
    \end{subfigure}
    \begin{subfigure}[b]{0.49\textwidth}
        \includegraphics[width=\textwidth]{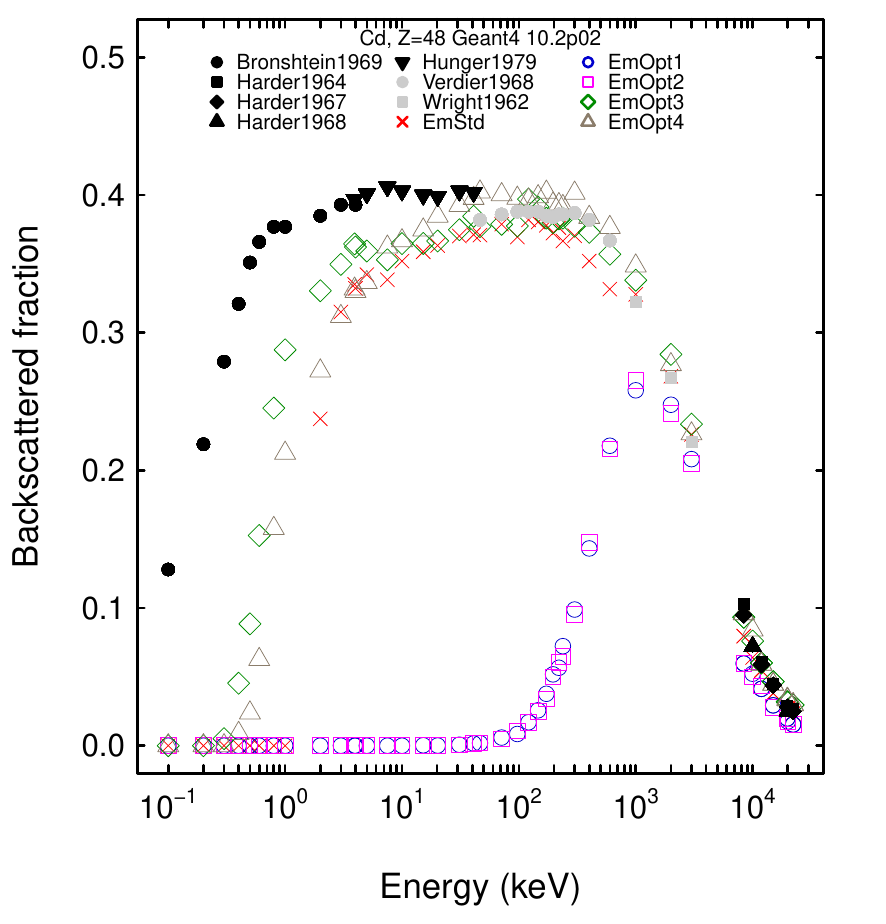}
        \label{fig_physListStd_v1022_48}
    \end{subfigure}
    \begin{subfigure}[b]{0.49\textwidth}
        \includegraphics[width=\textwidth]{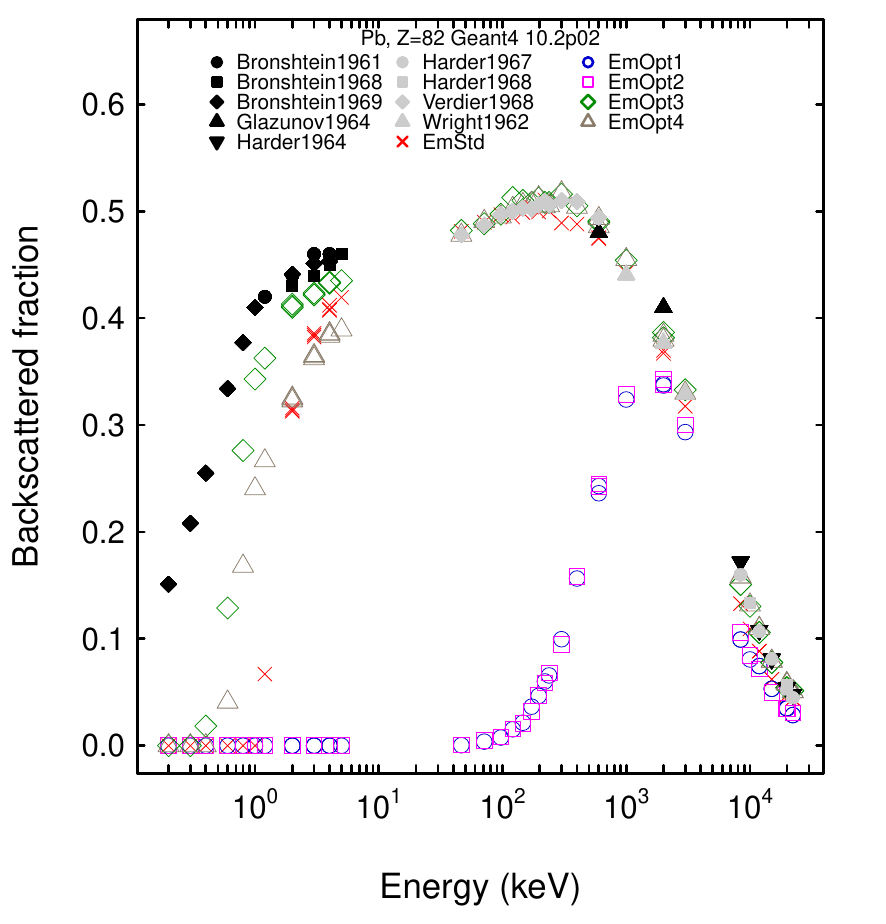}
        \label{fig_physListStd_v1022_82}
    \end{subfigure}
\caption{Measured and simulated fraction of backscattered electrons produced 
with different predefined electromagnetic PhysicsConstructors:
experimental data (black and grey filled markers); simulation with Geant4 10.2p02 
\textit{G4EmStandardPhysics} (EmStd, red crosses), 
\textit{G4EmStandardPhysics\_option1} (blue empty circles), 
\textit{G4EmStandardPhysics\_option2} (magenta empty squares), 
\textit{G4EmStandardPhysics\_option3} (green empty diamonds), 
\textit{G4EmStandardPhysics\_option4}  (brown empty triangles). 
The plots concern carbon, titanium, cadmium and lead targets.}
\label{fig_phOpt}
\end{figure*}

Differences across the simulation results associated with predefined
PhysicsConstructors and with respect to experimental data are qualitatively
visible in Figs.~\ref{fig_phGS} and \ref{fig_phOpt}; they have been
quantitatively investigated adopting a similar approach to the analysis of the
Goudsmit-Saunderson multiple scattering model.

For each energy range, the configuration based on the predefined
PhysicsConstructor that exhibits the highest efficiency is taken as a reference,
and its performance in terms of compatibility with experiment is compared with
the outcome of the Anderson-Darling test deriving from other configurations.
This analysis ascertains whether the predefined electromagnetic settings
distributed with Geant4 produce significantly different compatibility with
measurements; nevertheless, it is worhtwhile to note that
at lower energies the efficiencies obtained even with most efficient predefined
PhysicsContructors are substantially lower than those achieved above 100 keV.

The EmStd, EmSS and EmGS configurations,  based on the \textit{G4EmStandardPhysics},
\textit{G4EmStandardPhysicsSS} and \textit{G4EmStandardPhysicsGS} predefined
PhysicsConstructors, achieve the highest efficiency among the
simulations that use predefined PhysicsConstructors above 100 keV, in the 20-100
keV range and below 20 keV, respectively.
Their capability to reproduce experimental data is compared with that of other
simulation configurations; the resulting p-values are summarized in Table
\ref{tab_contph}.

As documented in Table \ref{tab_contph}, the hypothesis of equivalent
compatibility with experiment with respect to the most efficient
PhysicsConstructor is rejected for any other predefined PhysicsConstructors
below 20 keV, while it is rejected for all but \textit{G4StandardPhysicsWVI}
between 20 keV and 100 keV.
Above 100 keV, the hypothesis of compatibility with experiment equivalent to that 
achieved with \textit{G4EmStandardPhysics} is not rejected for the 
\textit{G4EmStandardPhysics\_option3}, \textit{G4EmStandardPhysics\_option4},
\textit{G4EmStandardPhysicsWVI} and \textit{G4EmStandardPhysicsSS} 
predefined PhysicsConstructors.


Equivalent compatibility with experiment with respect to the most efficient
PhysicsConstructor is also achieved with user-defined physics configurations: at
higher energy using single scattering or the WentzelVI multiple scattering
model, in the intermediate energy range with the UrbanBRF and WentzelBRFP
configurations, with some configuration options associated with the Goudsmit-Saunderson
multiple scattering model and with the Coulomb configuration, which uses the
\textit{G4eCoulombScatteringModel} single scattering model.

Neither \textit{G4EmStandardPhysics\_option3} nor \textit{G4Em\-Stan\-dard\-Physics\-\_option4},
which are recommended in \cite{g4appldevguide} ``for simulation with high
accuracy'',  ensure the highest achievable accuracy of the observable in
all Geant4-based simulation scenarios evaluated in this validation test.

The simulation configuration involving the \textit{G4Em\-Li\-ver\-mo\-re\-Phy\-sics}
PhysicsConstructor, which uses physics models encompassed in the Geant4 ``low
energy'' electromagnetic package, does not achieve equivalent compatibility with
experiment in the low energy range with respect to the more efficient configurations that use
\textit{G4EmStandardPhysicsGS} below 20 keV 
and \textit{G4EmStandardPhysicsSS} or \textit{G4EmStandardPhysicsWVI} between 20 and 100 keV.
The hypothesis of  equivalent compatibility with experiment is also rejected above 100 keV
with respect to simulations performed with \textit{G4EmStandardPhysics}.

The \textit{G4EmStandardPhysicsGS}  PhysicsConstructor achieves the highest efficiency 
among the predefined physics configurations examined in this test below 20 keV.
The results documented in section \ref{sec_GS} show that above 20 keV better
consistency with backscattering measurements can be achieved with configurations
using the Goudsmit-Saunderson multiple scattering model with settings other than
those implemented in the recommended \textit{G4EmStandardPhysicsGS}
PhysicsConstructor, as well as with
\textit{G4EmStandardPhysicsSS}, \textit{G4EmStandardPhysicsWVI} up to 100 keV 
and \textit{G4EmStandardPhysics} above 100 keV.

Some broad conclusions can be drawn from the outcome of this analysis.
As a general result, one can infer that no predefined PhysicsConstructor can
achieve compatibility with experiment  equivalent to the most efficient physics
configuration across the whole energy range covered by this validation test.
Additionally, predefined PhysicsConstructors explicitly labeled for ``high accuracy''
do not always ensure better compatibility with experimental data
than other available alternatives, nor does a PhysicsConstructor using models from the
Geant4 ``low energy'' electromagnetic package necessarily guarantee better
consistency with low energy measurements.

The results of this test offer guidance to simulation users to optimize the
selection of a predefined PhysicsConstructor appropriate to the characteristics 
of their experimental scenarios.

\subsubsection{Computational performance}

Qualitative indications about the computational performance associated with 
the predefined electromagnetic PhysicsConstructors can be derived from the plots in 
Figs. \ref{fig_time}  and \ref{fig_timeStd}.

Simulations with \textit{G4EmStandardPhysicsSS} and \textit{G4EmStandardPhysicsWVI}
are substantially slower than those with other predefined PhysicsConstructors;
Geant4 users concerned with limited availability of computational resources 
may want to consider user-defined physics configurations in the 20-100 keV
energy range, taking into account the results of compatibility with experiment 
listed in Table \ref{tab_contph} as guidance to investigate appropriate settings for 
their own experimental scenarios.

Simulations with \textit{G4EmStandardPhysicsGS} appear to be faster than with 
\textit{G4EmStandardPhysics} above a few keV, while simulations with 
\textit{G4EmLivermorePhysics} are generally slower; nevertheless, the 
difference in computational performance with respect to  \textit{G4EmStandardPhysics}
appear to be relatively small above a few MeV.
Similarly, the differences in computational speed observed with the generally
slower \textit{G4EmStandardPhysics\_option3},
\textit{G4\-Em\-Standard\-Physics\_option4} and with the generally faster
\textit{G4Em\-Standard\-Physics\_option1}, \textit{G4EmStandardPhysics\_option2}
options tend to decrease at higher energies.

\subsubsection{Energy Deposition}
\begin{figure} [tbhp]
\centerline{\includegraphics[angle=0,width=8.5cm]{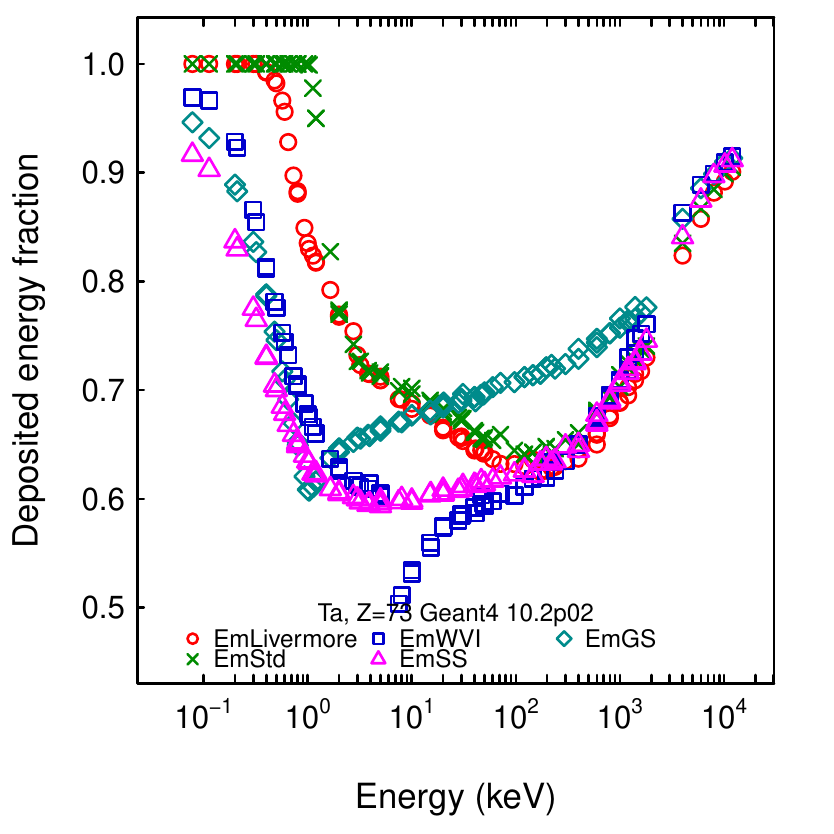}}
\caption{Fraction of the primary electron energy deposited in a tantalum target
as a function of the primary electron energy: 
G4EmStandardPhysics (EmStd, green crosses), 
G4EmStandardPhysicsGS (EmGS, turquoise downward triangles),
G4EmStandardPhysicsSS (EmSS, magenta upward triangles), 
G4EmStandardPhysicsWVI (EmWVI, blue squares) and 
G4EmLivermorePhysics (EmLiv, red circles). 
This plot has a qualitative character to illustrate the effects of different modeling
options on the simulation of energy deposition in a volume. }
\label{fig_edepPh}
\end{figure}

\begin{figure} [tbhp]
\centerline{\includegraphics[angle=0,width=8.5cm]{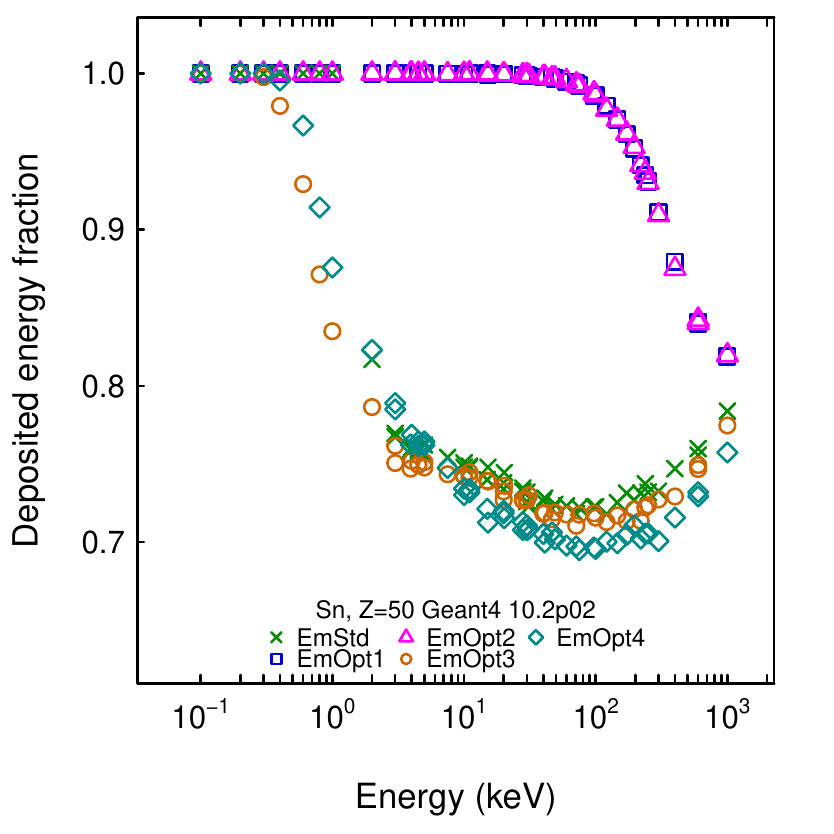}}
\caption{Fraction of the primary electron energy deposited in a tin target as a function 
of the primary electron energy: 
G4EmStandardPhysics (EmStd, green crosses), 
G4EmStandardPhysics\_option1 (EmOpt1, blue squares),
G4EmStandardPhysics\_option2 (EmOpt2, magenta upward triangles),  
G4EmStandardPhysics\_option3 (EmOpt3, turquoise downward triangles and 
G4EmStandardPhysics\_option4 (EmOpt4, brown diamonds). 
This plot has a qualitative character to illustrate the 
effects of different electron scattering options on the simulation of energy deposition in a volume.  }
\label{fig_edepPhStd}
\end{figure}

The simulation of the energy deposited in a volume is affected by the accuracy 
of the simulation of electron scattering.
In this respect, one should take into account that electrons can be either primary or 
secondary particles.

A detailed analysis of the effects of different options for electron scattering modeling
on the simulation of deposited energy is outside the scope of this paper;
moreover, the 
simple experimental model of backscattering experiments considered in this test
would not be adequate for in-depth studies of energy deposition simulation with Geant4.
Nevertheless, an example of the effects of different
predefined PhysicsConstructors is illustrated in Figs. \ref{fig_edepPh} and
\ref{fig_edepPhStd} for a qualitative appraisal.
The plots, based on Geant4 10.2p02, show the fraction of the primary electron energy that
is deposited in the targets modeled in this backscattering validation test:
differences are visible in the energy deposited in the target associated
with the various predefined physics options.

\section{Epistemological Considerations}


Some epistemological guidelines should be taken into account in the
appraisal of the results documented in this paper, as well as in similar
contexts of validation tests concerning observables produced in use cases of
Monte Carlo simulation codes.

Although the observable considered in this validation test is sensitive to
electron scattering modeling, its simulated outcome is the result of all the
code involved in the simulation, including direct and indirect dependencies,
and of the computational environment where the simulation is produced.
The simulation configurations studied in this paper are identified for
convenience through the electron scattering modeling they involve, nevertheless
possible effects of other parts of the simulation code on shaping the observable
subject to validation should not be neglected.
This limitation of inference is common to any observable produced as a result of
a Monte Carlo simulation system.
In this respect, a validation test specific to a simulation use case is
epistemologically distinct from the context of the validation of parameters used
in Monte Carlo codes, e.g. cross sections, which can be compared with
experimental measurements independently from the Monte Carlo software systems
where they are used.

It is worthwhile to remind the reader that all the results associated with the
simulation configurations considered in this paper concern the test of a specific
observable, i.e. the fraction of backscattered electrons. 
Caution should be exercised in extrapolating them to assess the reliability of
other simulated observables not subject to validation in this paper, or to other
physically different environments, e.g. at lower or higher energies \cite{trucano_predict}.


Known concerns related to the role of induction in establishing scientific
knowledge \cite{popperLSD}, which question the
foundation of general statements about the validity or the accuracy of
simulation models, should be taken into account.
A correct approach consists of documenting the context in which
the behaviour of models has been empirically quantified.


\section{Conclusion}

The results collected in this paper document validation tests of Geant4-based
simulation of electron backscattering with a wide variety of physics modelling
options available in the Geant4 toolkit.
Statistical data analysis methods allow quantitative and objective appraisal of
the capabilities of different physics configurations to produce simulation
results compatible with experimental measurements.

Significant differences are observed across the set of physics options subject to test,
regarding their ability to generate results consistent with the measured fraction 
of backscattered electrons.
No single physics modeling configuration is capable of producing optimal
results over the whole energy range covered by the validation test. 
The detailed validation analysis summarized in this paper provides guidance to help experimental
users identify, among the many possible options available in the toolkit,
those that most effectively address the requirements specific to their own
experimental scenarios.
Comparative evaluations of the computational performance of the simulation
configurations, documented along with the results of validation tests, provide
complementary information, although at a qualitative level, to guide the
selection.

The results collected in this paper show the benefit of validation tests of
simulation modeling options, whose capabilities are quantified with respect
to large experimental data samples through statistical analysis methods.
Their inclusion in the software development process of simulation toolkits 
is beneficial to the experimental
community, both to support the optimization of new models in the course of their
development and to provide users with an objective characterization of the
behaviour of the software released for scientific applications.




%
\section*{Acknowledgment}

The authors thank Alessandro Brunengo and Mirko Corosu (INFN Genova Computing
Service) for helpful support to the simulation production, Gabriele Cosmo, Tatiana Nikitina 
and Mauro Tacconi for useful information regarding Geant4 code they
developed or managed, and Anita Hollier for proofreading the manuscript.
The support of the CERN Library has been essential to collect the experimental data 
used in the validation test.


\end{document}